\documentclass[twocolumn,preprintnumbers,amsmath,amssymb,floatfix]{revtex4}
\usepackage{graphicx}
\usepackage{dcolumn}
\usepackage{bm}
\begin{document}
\ifx\href\undefined\else\hypersetup{linktocpage=true}\fi 
\title{Metal-insulator transitions and the effects of electron-electron
interactions in two-dimensional electron systems}
\author{A.~A. Shashkin}
\affiliation{Institute of Solid State Physics, Chernogolovka, Moscow
District 142432, Russia}
\begin{abstract}
Experimental results on the metal-insulator transitions and the
anomalous properties of strongly interacting two-dimensional electron
systems are reviewed and critically analyzed. Special attention is
given to recent results for the strongly enhanced spin susceptibility
and effective mass in low-disordered silicon MOSFETs.
\end{abstract}
\pacs{71.30.+h,73.40.Qv}
\maketitle
\tableofcontents

\section{INTRODUCTION}

Two-dimensional (2D) electron systems are realized when the electrons
are free to move in a plane but their motion perpendicular to the
plane is quantized in a confining potential well. At low electron
densities in such systems, the strongly-interacting limit is reached
because the kinetic energy is overwhelmed by the energy of
electron-electron interactions. The interaction strength is
characterized by the ratio between the Coulomb energy and the Fermi
energy, $r_s^*=E_{ee}/E_F$. Assuming that the effective electron mass
is equal to the band mass, the interaction parameter $r_s^*$ in the
single-valley case reduces to the Wigner-Seitz radius, $r_s=1/(\pi
n_s)^{1/2}a_B$ and therefore increases as the electron density,
$n_s$, decreases (here $a_B$ is the Bohr radius in semiconductor).
Possible candidates for the ground state of the system include Wigner
crystal characterized by spatial and spin ordering \cite{wigner34},
ferromagnetic Fermi liquid with spontaneous spin ordering
\cite{stoner46}, paramagnetic Fermi liquid \cite{landau57}, etc. In
the strongly-interacting limit ($r_s\gg1$), no analytical theory has
been developed to date. According to numeric simulations
\cite{tanatar89}, Wigner crystallization is expected in a very dilute
regime, when $r_s$ reaches approximately 35. The refined numeric
simulations \cite{attaccalite02} have predicted that prior to the
crystallization, in the range of the interaction parameter
$25\lesssim r_s\lesssim35$, the ground state of the system is a
strongly correlated ferromagnetic Fermi liquid. At higher electron
densities, $r_s\sim1$, the electron liquid is expected to be
paramagnetic, with the effective mass, $m$, and Land\'e $g$ factor
renormalized by interactions. Apart from the ferromagnetic Fermi
liquid, other intermediate phases between the Wigner crystal and the
paramagnetic Fermi liquid may also exist.

In real 2D electron systems, the inherent disorder leads to a drastic
change of the above picture, which significantly complicates the
problem. According to the scaling theory of localization
\cite{abrahams79}, all electrons in a disordered infinite
noninteracting 2D system become localized at zero temperature and
zero magnetic field. At finite temperatures, regimes of strong and
weak localizations are distinguished: (i)~if the conductivity of the
2D electron layer is activated, the resistivity diverges
exponentially as $T\rightarrow0$; and (ii)~in the opposite limit of
weak localization the resistivity increases logarithmically with
decreasing temperature --- an effect originating from the increased
probability of electron scattering from impurities back to the
starting point. Interestingly, the incorporation of weak interactions
($r_s<1$) between the electrons promotes the localization
\cite{altshuler80}. However, for weak disorder and $r_s\gtrsim1$ a
possible metallic ground state was predicted
\cite{finkelstein83,finkelstein84,castellani84}.

\begin{figure}
\scalebox{0.4}{\includegraphics[clip]{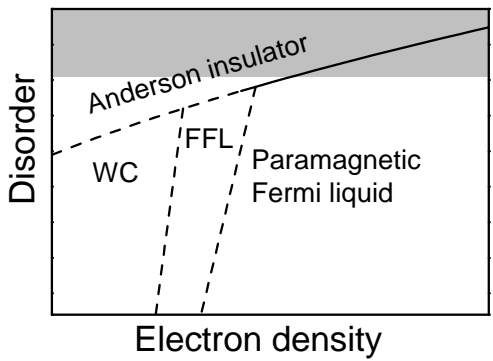}}
\caption{\label{diagram} Schematic phase diagram in disorder vs $n_s$
plane. The Wigner crystal (WC) regime is preceded by the
ferromagnetic Fermi liquid (FFL) \cite{attaccalite02}. The
high-disorder region is shaded.}
\end{figure}

In view of the competition between the interactions and disorder,
high- and low-disorder limits can be considered. Since many of the
experimental groups have made little distinction between these, there
has been confusion about interpretation of the data. In
highly-disordered electron systems, the range of low densities is not
accessible as the strong (Anderson) localization sets in (see
Fig.~\ref{diagram}). At higher densities, a logarithmic-in-$T$
correction to the resistivity was observed in numerous experiments
(see, e.g., Refs.~\cite{dolan79,bishop80,uren80}), providing support
for the weak localization theory. Apparently, extrapolation of the
weak corrections to $T=0$ is not justified and, therefore, those
studies cannot serve as confirmation of the scaling theory. The
latter is remarkable by the principal impossibility of experimental
verification because all experiments are performed in samples with
finite dimensions at finite temperatures. The question whether or not
the scaling theory works is essentially a matter of belief.

The case of low-disordered electron systems is much more interesting.
Low electron densities corresponding to the strongly-interacting
limit become accessible. Experimental results on the metal-insulator
phase diagram in perpendicular magnetic fields revealed a close
similarity between the insulating phase at low densities and the
quantum Hall states. Thus, they excluded the formation of a pinned
Wigner crystal in available samples, but supported the existence of a
metallic state in zero field
\cite{shashkin93,shashkin94a,shashkin94b}. As the magnetic field is
decreased, the extended states in the Landau levels were observed to
float up in energy relative to the Landau level centers and merge to
form a metallic state in the $B=0$ limit. This contradicts the
theoretical scenario that in the limit of zero magnetic field the
extended states should float up indefinitely in energy
\cite{khmelnitskii84,laughlin84} leading to an insulating ground
state. The metallic state was found to be remarkable by the strong
drop of resistivity with decreasing temperature
\cite{kravchenko94a,kravchenko95a,kravchenko96}. Although the origin
of the effect was attributed to strong electron-electron
interactions, the underlying physics remained unclear until recently.

A breakthrough in understanding this topic occurred in the past four
years. After a strongly enhanced ratio $gm$ of the spin and the
cyclotron splittings was found at low $n_s$ in silicon MOSFETs
\cite{kravchenko00a}, it became clear that the system behaves well
beyond the weakly interacting Fermi liquid. It was reported that the
magnetic field required to produce complete spin polarization,
$B_c\propto n_s/gm$, tends to vanish at a finite electron density
$n_\chi\approx 8\times 10^{10}$~cm$^{-2}$, which is close to the
critical density, $n_c$, for the metal-insulator transition in this
electron system \cite{shashkin01a,vitkalov01a,shashkin02a}. These
findings point to a sharp increase of the spin susceptibility,
$\chi\propto gm$, and possible ferromagnetic instability in dilute
silicon MOSFETs. In very dilute GaAs/AlGaAs heterostructures, a
similar behavior has been observed in both 2D hole and 2D electron
systems \cite{gao02,zhu03}. Recently, experimental results have
indicated that in silicon MOSFETs it is the effective mass, rather
than the $g$ factor, that sharply increases at low electron densities
\cite{shashkin02b}. They have also indicated that the anomalous rise
of the resistivity with temperature is related to the increased mass.
The magnitude of the mass does not depend on the degree of spin
polarization, which points to a spin-independent origin of the
effective mass enhancement \cite{shashkin03a,shashkin03b}. It is
interesting that the observed effects are more pronounced in silicon
MOSFETs compared to GaAs/AlGaAs heterostructures, although the
fractional quantum Hall effect, which is usually attributed to
electron-electron interactions, has not been reliably established in
silicon MOSFETs.

The fact that $n_\chi$ is close to the critical density $n_c$
indicates that the metal-insulator transition in silicon samples with
very low disorder potential is a property of a clean 2D system and is
driven by interactions \cite{shashkin01a}. This is qualitatively
different from a localization-driven transition in more-disordered
samples that occurs at appreciably higher densities than $n_\chi$
which are also dependent on disorder strength. In this review,
attention is focused on results obtained in the clean regime.

There are several good reviews on the topic in question (see, e.g.,
Refs.~\cite{abrahams01,kravchenko04}). However, they have either
become outdated or do not express criticism toward the items of
general belief, e.g., the scaling theory. Below, I describe the main
experimental results and give a broad view on the metal-insulator
transition and anomalous properties of 2D electron systems at low
densities.

\section{METAL-INSULATOR PHASE DIAGRAMS IN A MAGNETIC FIELD}

Metal-insulator transitions in perpendicular magnetic fields
attracted a great deal of interest in the past decade. The
experimental activity was strongly stimulated by theoretical
predictions that Wigner crystallization is promoted in the presence
of a magnetic field
\cite{lozovik75,tsukada77,maki83,lam84,levesque84}. The insulating
phase at low electron densities was mainly studied whose origin was
attributed to possible formation of the Wigner crystal
\cite{pudalov90,kravchenko91,iorio92,pudalov93a,willett88,goldman88,andrei88,willett89,jiang90,goldman90,williams91,jiang91,santos92a,santos92b,manoharan94}.
However, the latter has been precluded in studies of the
metal-insulator phase diagram including quantum Hall states, which
show a close similarity of all insulating phases in available samples
\cite{shashkin93,shashkin94a,shashkin94b}. It is interesting that
there are some firmly-established experimental results which have not
attracted much of the theorists' attention. These include
(i)~oscillations of the metal-insulator phase boundary as a function
of perpendicular magnetic field; and (ii)~finite bandwidth of the
extended states in the Landau levels.

\subsection{Floating-up of the extended states in perpendicular magnetic fields}
\label{floating-up}

\subsubsection{First observation}

\begin{figure}
\scalebox{0.55}{\includegraphics[clip]{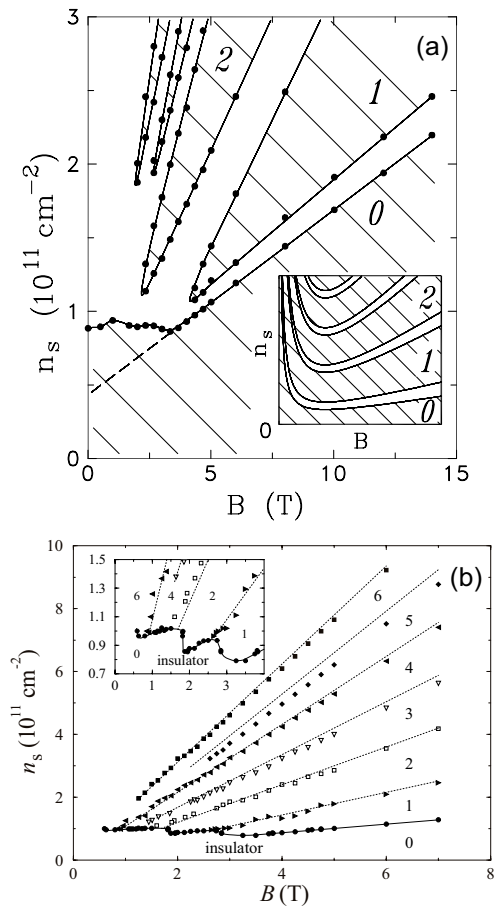}}
\caption{\label{floating} (a)~Metal-insulator phase diagram in a
low-disordered 2D electron system in silicon MOSFETs obtained using a
cutoff criterion, $\sigma_{xx}=e^2/20h$, at a temperature $\approx
25$~mK. The dimensionless $\sigma_{xy}h/e^2$ in different insulating
phases is indicated. The slope of the dashed line is close to
$e/2hc$. A sketch of the expected phase diagram is displayed in the
inset. From Ref.~\cite{shashkin93}. (b)~The map of extended states
determined by maxima in $\sigma_{xx}$ in a low-disordered silicon
MOSFET. Numbers show $\sigma_{xy}$ in units of $e^2/h$. Adopted from
Ref.~\cite{kravchenko95b}.}
\end{figure}

The scaling theory of localization was challenged by the quantum Hall
effect (quantization of the Hall resistivity, $\rho_{xy}=h/\nu e^2$,
at integer filling factor $\nu$ accompanied with vanishing
longitudinal resistivity, $\rho_{xx}$) \cite{klitzing80} which
implies the existence of extended states in the Landau levels (see
section~\ref{edge}). To reconcile these two, it was theoretically
predicted almost immediately that the extended states in the Landau
levels cannot disappear discontinuously with decreasing magnetic
field but must float up indefinitely in energy in the limit of $B=0$
\cite{khmelnitskii84,laughlin84}. The expected phase diagram is shown
in the inset to Fig.~\ref{floating}(a). An equivalent diagram plotted
in disorder versus inverse filling factor ($1/\nu=eB/hcn_s$) plane is
known as the global phase diagram for the quantum Hall effect
\cite{kivelson92}. As long as no merging of the extended states was
considered to occur, their piercing of the Fermi level was predicted
to cause quantization of the Hall conductivity in weak magnetic
fields \cite{khmelnitskii92,huckestein00}.

The first attempt \cite{shashkin93} to experimentally determine the
metal-insulator phase diagram at low temperatures in low-disordered
silicon MOSFETs already revealed discrepancies with the theory (see
Fig.~\ref{floating}(a)). In that paper, a somewhat arbitrary
criterion for the longitudinal conductivity, $\sigma_{xx}=e^2/20h$,
was used to map out the phase boundary that corresponds to the
Anderson transition to the regime of strong localization. However,
firstly, the phase boundary was shown to be insensitive to the choice
of the cutoff value (see, e.g., Ref.~\cite{dolgopolov92b}), and
secondly, that particular cutoff value is consistent with the results
obtained for quantum Hall states by a vanishing activation energy
combined with a vanishing nonlinearity of current-voltage
characteristics when extrapolated from the insulating phase
\cite{shashkin94a} (note that for the lowest-density phase boundary,
a lower value $\sigma_{xx}^{-1}\approx 100$~kOhm at a temperature
$\approx 25$~mK follows from the latter method). The metallic phase
surrounds each insulating phase as characterized by the dimensionless
Hall conductivity, $\sigma_{xy}h/e^2$, that counts the number of
quantum levels below the Fermi level (in bivalley (100)-silicon
MOSFETs, spin and valley degeneracies of the Landau level should be
taken into account). This indicates that the extended states indeed
do not disappear discontinuously. However, with decreasing magnetic
field they float up in energy relative to the Landau level centers
and merge forming a metallic state in the limit of $B=0$ (see
sections~\ref{similarity} and \ref{true}). Besides, the phase
boundary at low electron densities oscillates as a function of $B$
with minima corresponding to integer filling factors. The phase
boundary oscillations manifest themselves in that at electron
densities near the $B=0$ metal-insulator transition, the
magnetoresistance oscillates with an amplitude that diverges as
$T\rightarrow0$ \cite{pudalov90}; the regions in which the
magnetoresistance diverges are referred to as reentrant insulating
phase (see section~\ref{similarity}).

\subsubsection{Other methods and 2D carrier systems}

\begin{figure}
\scalebox{0.54}{\includegraphics[clip]{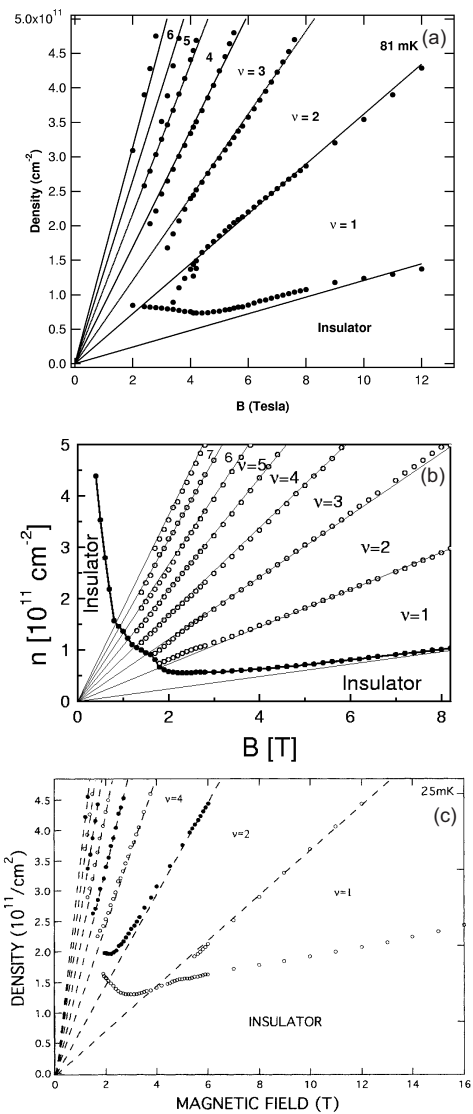}}
\caption{\label{dultz} (a)~A map of the extended states for a
highly-disordered 2D hole system in GaAs/AlGaAs heterostructures.
Each data point represents a distinct peak in $\sigma_{xx}$ or a
temperature independent crossing point in $\rho_{xx}$. Numbers show
the value of $\sigma_{xy}h/e^2$. Adopted from Ref.~\cite{dultz98}.
(b)~A map of the extended states for a highly-disordered 2D hole
system in a Ge/SiGe quantum well. The open circles represent maxima
in $\rho_{xx}$ and/or $d\rho_{xy}/dB$. The solid circles correspond
to crossing points of $\rho_{xx}$ at different temperatures. Numbers
show the value of $\sigma_{xy}h/e^2$. Adopted from
Ref.~\cite{hilke00}. (c)~Behavior of the extended states determined
by maxima in $\sigma_{xx}$ in a strongly-disordered 2D electron
system in GaAs/AlGaAs heterostructures. Numbers show $\sigma_{xy}$ in
units of $e^2/h$. Adopted from Ref.~\cite{glozman95}.}
\end{figure}

The topology of the observed metal-insulator phase diagram ---
merging the extended states and, hence, the presence of direct
transitions between the insulating phase with $\sigma_{xy}=0$ and
quantum Hall phases with $\sigma_{xy}h/e^2>1$ --- is robust being
insensitive to the method for spotting the phase boundary
\cite{shashkin94a,kravchenko95b} and to the choice of 2D carrier
system \cite{dultz98,hilke00}. It was verified using a criterion of
vanishing activation energy and vanishing nonlinearity of
current-voltage characteristics as extrapolated from the insulating
phase, which allows more accurate determination of the Anderson
transition \cite{shashkin94a}. Kravchenko et al. \cite{kravchenko95b}
also applied for similar silicon MOSFETs a method that had been
suggested in Ref.~\cite{glozman95}. They studied extended states by
tracing maxima in the longitudinal conductivity in the ($B,n_s$)
plane (see Fig.~\ref{floating}(b)) and found a good agreement with
the aforementioned results. A similar merging of at least the two
lowest extended states was observed in more-disordered 2D hole
systems in a GaAs/AlGaAs heterostructure \cite{dultz98} (see
Fig.~\ref{dultz}(a)) and in a Ge/SiGe quantum well \cite{hilke00}
(see Fig.~\ref{dultz}(b)). In the former case the extended states
were determined by peaks in $\sigma_{xx}$ or temperature-independent
crossing points in $\rho_{xx}$; in the latter they were associated
either with maxima in $\rho_{xx}$ and/or $d\rho_{xy}/dB$, or with
crossing points of $\rho_{xx}$ at different temperatures. It is
noteworthy that a bad combination of the criterion for determining
the phase boundary and the 2D carrier system under study may lead to
a failure of mapping out the phase diagram down to relatively weak
magnetic fields. In Ref.~\cite{glozman95}, extended states were
studied by measuring maxima in the longitudinal conductivity in the
($B,n_s$) plane for the strongly-disordered 2D electron system in
GaAs/AlGaAs heterostructures (see Fig.~\ref{dultz}(c)). Because of
strong damping of the Shubnikov-de~Haas oscillations in low magnetic
fields, the desired region on the phase diagram below 2~T was not
accessible in that experiment. This invalidates the claim of Glozman
et al. \cite{glozman95} that the extended states do not merge
\cite{shashkin95}. The behavior of the lowest extended state in
Fig.~\ref{dultz}(c), which Glozman et al. \cite{glozman95} claim to
float up above the Fermi level as $B\rightarrow0$, simply reflects
the occurrence of a phase boundary oscillation minimum at filling
factor $\nu=2$, similar both to the minimum at $\nu=1$ in
Fig.~\ref{dultz}(a) and to the case of silicon MOSFETs
(Fig.~\ref{floating}). Such a minimum manifests itself in that there
exists a minimum in $\rho_{xx}$ at integer $\nu\ge1$ that is
straddled by the insulating phase
\cite{pudalov90,pudalov93b,jiang93,shahar95a,hilke97,sakr01}. To this
end, all available data for the metal-insulator phase diagrams agree
well with each other, except those in the vicinity of $B=0$. In weak
magnetic fields, experimental results obtained in 2D electron systems
with high disorder are not method-independent. Glozman et al.
\cite{glozman95} found that the cutoff criterion yields basically a
flat phase boundary towards $B=0$, which is in agreement with the
data for silicon MOSFETs (Fig.~\ref{floating}(a)). On the contrary,
Hilke et al. \cite{hilke00} employed the method based on temperature
dependences of $\rho_{xx}$ and obtained a turn up on the phase
boundary in Fig.~\ref{dultz}(b). Note that the validity of the data
for the lowest extended state at magnetic fields $\lesssim1.5$~T in
Fig.~\ref{dultz}(b) is questionable because the weak temperature
dependences of $\rho_{xx}$ as analyzed by Hilke et al. \cite{hilke00}
cannot be related to either an insulator or a metal. The same applies
to similar temperature dependences observed, e.g., in
Refs.~\cite{wang94,hughes94,song97,lee98,hilke98,hilke99,hanein99}.

\subsubsection{Weak-field regime}

As a matter of fact, the weak-field problem --- whether or not there
is an indefinite rise of the phase boundary as $B\rightarrow0$ --- is
a problem of the existence of a metal-insulator transition at $B=0$
and $T=0$. In dilute 2D electron systems with low enough disorder,
the resistivity, $\rho$, strongly drops with lowering temperature
providing an independent way of facing the issue. Given strong
temperature dependences of $\rho$, those with $d\rho/dT>0$
($d\rho/dT<0$) can be associated with a metallic (insulating) phase
\cite{kravchenko94a,kravchenko95a,kravchenko96,popovic97,coleridge97}.
If extrapolation of the temperature dependences of $\rho$ to $T=0$ is
valid, the curve with $d\rho/dT=0$ should correspond to the
metal-insulator transition. The fact that this method and the one,
based on a vanishing activation energy combined with a vanishing
nonlinearity of current-voltage curves when extrapolated from the
insulating phase, give equivalent results strongly supports the
existence of a true metal-insulator transition in zero magnetic field
\cite{shashkin01b} (see section~\ref{true}). As long as in
more-disordered 2D carrier systems the metallic ($d\rho/dT>0$)
behavior is suppressed (see, e.g.,
Refs.~\cite{papadakis98,hanein98,simmons98,mills99,yoon99,simmons00,noh03,pudalov01})
or disappears entirely, it is definitely incorrect to extrapolate
those weak temperature dependences of $\rho$ to $T=0$ with the aim to
distinguish between insulator and metal. Once one of the two methods
fails, it remains to be seen how to verify the conclusion as inferred
from the other method. This makes uncertain the existence of a true
$B=0$ metal-insulator transition in 2D electron systems with high
disorder.

\subsubsection{Phase boundary oscillations}

\begin{figure}
\scalebox{0.55}{\includegraphics[clip]{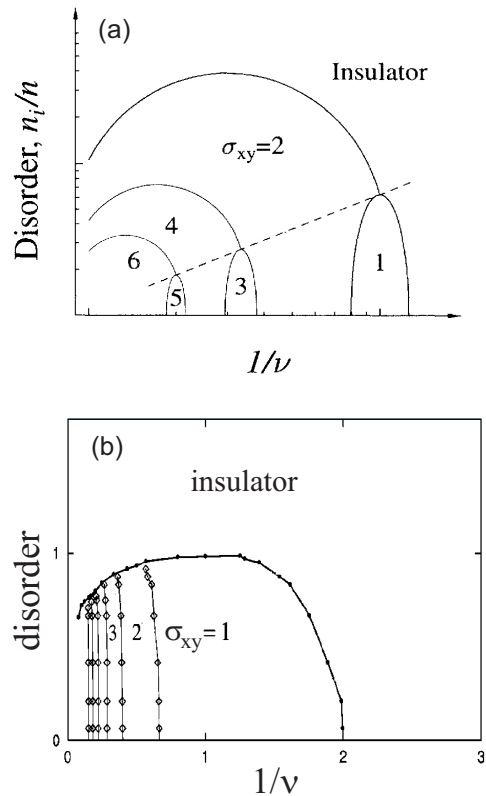}}
\caption{\label{fogler} (a)~A sketch of the modified global phase
diagram for the quantum Hall effect as expected from the mean-field
approximation neglecting the Zeeman energy. The dashed line
corresponds to the collapse of the exchange-enhanced spin splitting.
Adopted from Ref.~\cite{fogler95}. (b)~Numerical results for the
phase diagram within a tight-binding model. Adopted from
Ref.~\cite{sheng00}.}
\end{figure}

The next important point is the oscillating behavior of the phase
boundary that restricts the insulating phase with $\sigma_{xy}=0$
(see, e.g., Fig.~\ref{floating}). It is worth noting that the
oscillations persist down to the magnetic fields corresponding to the
fillings of higher Landau levels, as indicated also by
magnetoresistance oscillations \cite{pudalov90,pudalov93b,sakr01}.
The oscillation period includes the following elements. With
decreasing magnetic field the lowest extended states follow the
Landau level, float up in energy relative to its center, and merge
with extended states in the next quantum level. The last element was
absent in the original considerations
\cite{khmelnitskii84,laughlin84,kivelson92,khmelnitskii92} leading to
discrepancies between experiment and theory. Recently, theoretical
efforts have been concentrated on modifications of the global phase
diagram for the quantum Hall effect to reach topological
compatibility with the observed metal-insulator phase diagram. It has
been predicted that the spin-up and spin-down extended states in the
Landau level should merge \cite{fogler95,tikofsky00} (see
Fig.~\ref{fogler}(a)). However, in view of lowest extended states,
the topology of the phase diagram changes for the lowest Landau level
only; besides, they do not float up before merging. It has been
verified that shifts of the extended states from the Landau level
centers that are caused by disorder-induced mixing of the Landau
levels are small
\cite{ando84,shahbazyan95,kagalovsky95,gramada96,haldane97,fogler98}.
Within tight-binding models, an indication had been first obtained
that the extended states disappear \cite{liu96,xie96,hatsugai99},
which caused some criticism toward the relevance of such a lattice
model to the continuum system \cite{yang96,yang99}. After that,
floating up of the extended states without merging has been found in
studies \cite{koschny01,pereira02,koschny03}. In contrast, Sheng and
Weng \cite{sheng97,sheng00} have obtained a merging of the extended
states although without an oscillating behavior of the lowest
extended states (see Fig.~\ref{fogler}(b)). As for now, the effect of
the phase boundary oscillations is still far from being described
theoretically.

Concluding this section, I would like to make a couple of remarks on
alternative ways for determining the metal-insulator phase boundary.
An attempt was made to spot the phase boundary in the limit of $B=0$
using a criterion of $\sigma_{xy}=e^2/2h$ \cite{okamoto95}. However,
that particular value of $\sigma_{xy}$ has no special meaning as
$B\rightarrow0$. An idea was expressed to relate the minimum in the
inverse compressibility to the metal-insulator transition
\cite{dultz00}. However, it has been recently shown that in zero
magnetic field, this minimum lies at carrier densities well above the
critical density for the percolation metal-insulator transition
\cite{fogler04}. Particularly, in highly-disordered 2D carrier
systems, its position may be close to that of the crossing point of
the resistivity at different temperatures \cite{dultz00} which
formally yields overestimated densities for the metal-insulator
transition because of suppression of the metallic behavior (see
section~\ref{true}).

\subsection{Similarity of the insulating phase and quantum Hall phases}
\label{similarity}

\subsubsection{Method for comparison and consequences}

About a decade ago, main attention was paid to the insulating phase
at low electron densities as a possible candidate for the Wigner
crystal. It was argued that its aforementioned reentrant behavior is
a consequence of the competition between the quantum Hall effect and
the pinned Wigner crystal \cite{pudalov90,kravchenko91}. Another
certain argument was strongly nonlinear current-voltage
characteristics in the insulating phase which were attributed to
depinning of the Wigner crystal \cite{iorio92,pudalov93a}. Similar
features of the insulating phase in a 2D electron (near $\nu=1/5$)
\cite{willett88,goldman88,andrei88,willett89,jiang90,goldman90,williams91,jiang91}
and 2D hole (near $\nu=1/3$) \cite{santos92a,santos92b,manoharan94}
systems in GaAs/AlGaAs heterostructures with relatively low disorder
were also attributed to a pinned Wigner crystal which is interrupted
by the fractional quantum Hall state. An alternative scenario was
discussed in terms of percolation metal-insulator transition
\cite{dolgopolov92b,dolgopolov92a,dolgopolov92c}. To distinguish
between the two scenarios, the behavior of activation energy and
current-voltage characteristics in the insulating phase was studied
and compared to that in quantum Hall phases
\cite{shashkin94a,shashkin94b,dolgopolov95}.

\begin{figure}
\scalebox{0.46}{\includegraphics[clip]{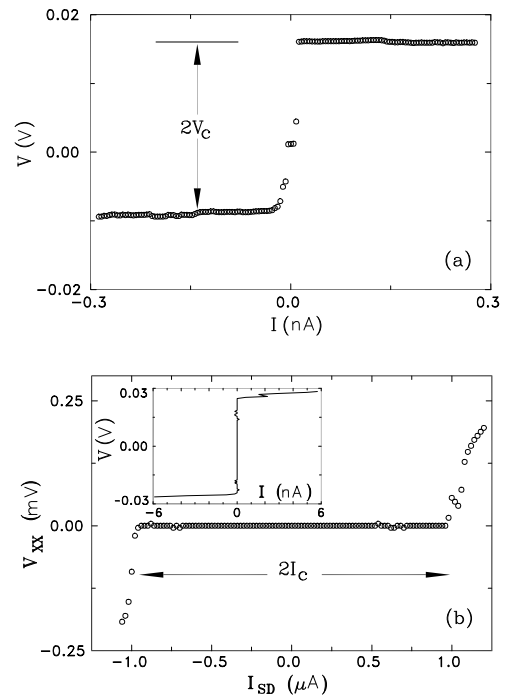}}
\caption{\label{iv} Current-voltage characteristics in a
low-disordered silicon MOSFET in $B=12$~T at $T\approx 25$~mK for the
low-density insulating phase at $n_s=1.74\times 10^{11}$~cm$^{-2}$
(a) and the insulating phase with $\sigma_{xy}h/e^2=1$ at
$n_s=2.83\times 10^{11}$~cm$^{-2}$ (b). In (b) the measured breakdown
dependence $V_{xx}(I_{sd})$ is converted into current-voltage
characteristics (inset). From Ref.~\cite{shashkin94a}.}
\end{figure}

\begin{figure}
\scalebox{0.4}{\includegraphics[clip]{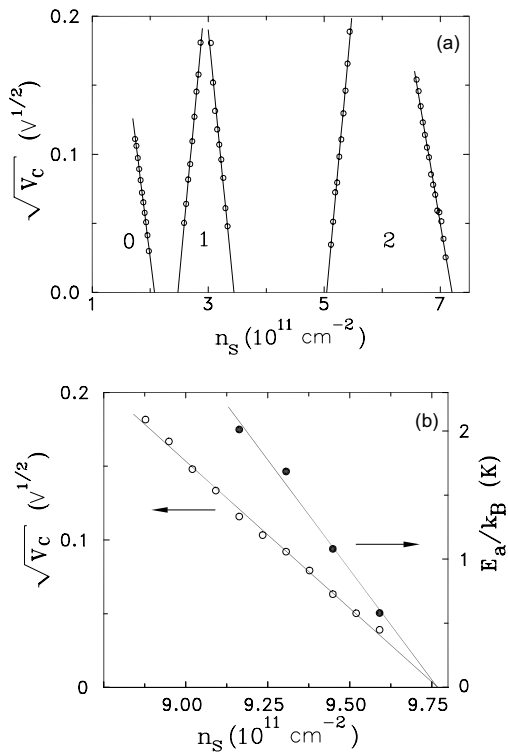}}
\caption{\label{Vc} (a)~Square root of the critical voltage as a
function of electron density at the phase boundaries corresponding to
$\sigma_{xy}h/e^2=0$, 1, and 2 in $B=12$~T for a low-disordered 2D
electron system in silicon MOSFETs. (b)~Behavior of the critical
voltage and the activation energy near the phase boundary in
$B=16$~T. From Ref.~\cite{shashkin94a}.}
\end{figure}

In contrast to the low-density insulating phase, the way of
determining the current-voltage characteristics of the quantum Hall
phases is different for Corbino and Hall bar geometries. In the
former the dissipationless Hall current does not contribute to the
dissipative current that is proportional to $\sigma_{xx}$, allowing
straightforward measurements of current-voltage curves for all
insulating phases. In the latter the two current channels are
connected through edge channels (see section~\ref{edge}), and
current-voltage characteristics correspond to quantum-Hall-effect
breakdown curves. The dissipative backscattering current, $I$, that
flows between opposite edge channels is balanced by the Hall current
in the filled Landau levels associated with the longitudinal voltage,
$V_{xx}$. As long as $\sigma_{xx}\ll\sigma_{xy}$, the quantized value
of $\sigma_{xy}$ is a factor that allows determination of
$I=\sigma_{xy}V_{xx}$ and the Hall voltage, $V=I_{sd}/\sigma_{xy}$,
from the experimental breakdown dependence of $V_{xx}$ on
source-drain current, $I_{sd}$. The dependence $V(I)$ is a
current-voltage characteristic, which is equivalent to the case of
Corbino geometry \cite{shashkin94a} (see Fig.~\ref{iv}). Not only are
the current-voltage curves similar for all insulating phases, but
they behave identically near the metal-insulator phase boundaries
(see Fig.~\ref{Vc}(a)). The dependence of the critical voltage,
$V_c$, on the distance from the phase boundary is close to a
parabolic law \cite{dolgopolov92b,pudalov93a}. The phase boundary
position determined by a vanishing $V_c$ is practically coincident
with that determined by a vanishing activation energy, $E_a$, of
electrons from the Fermi level $E_F$ to the mobility edge, $E_c$ (see
Fig.~\ref{Vc}(b)). The value $E_a$ is determined from the temperature
dependence of the conduction in the linear interval of
current-voltage curves, which is activated at not too low
temperatures \cite{adkins76}; note that it transforms into variable
range hopping as $T\rightarrow0$ (see below). The activation energy
changes linearly with the distance from the phase boundary reflecting
constancy of the thermodynamic density of states near the transition
point (see also section~\ref{true}). The threshold behavior of the
current-voltage characteristics is caused by the breakdown in the
insulating phases. The breakdown occurs when the localized electrons
at the Fermi level gain enough energy to reach the mobility edge in
an electric field, $V_c/d$, over a distance given by the localization
length, $L$ \cite{shashkin94a,polyakov93}:
\begin{equation}eV_cL/d=|E_c-E_F|,\label{break}\end{equation}
where $d$ is the corresponding sample dimension. The values $E_a$ and
$V_c$ are related through the localization length which is
temperature independent and diverges near the transition as
$L(E_F)\propto |E_c-E_F|^{-s}$ with exponent $s$ close to unity, in
agreement with the theoretical value $s=4/3$ in classical percolation
problem \cite{shklovskii84}. The value of the localization length is
practically the same near all metal-insulator phase boundaries, which
indicates that even quantitatively, all insulating phases are very
similar. Note that since the localization length in Eq.~(\ref{break})
is small compared to the sample sizes, the phase boundary position
determined by the diverging localization length refers to an infinite
2D system. As inferred from the vanishing of both $E_a$ and $V_c$ at
the same point (see Fig.~\ref{Vc}(b)), possible shifts of the
mobility threshold due to finite sample dimensions are small, which
justifies extrapolations to the limit of $L\rightarrow\infty$.

\begin{figure}
\scalebox{0.33}{\includegraphics[clip]{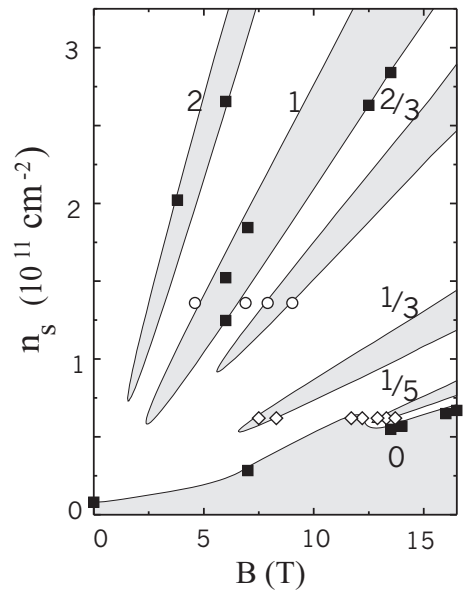}}
\caption{\label{gaas} Metal-insulator phase diagram in a relatively
low-disordered 2D electron system in GaAs/AlGaAs heterostructures.
The points corresponding to phase boundaries are obtained on two
samples using a criterion of vanishing activation energy and
vanishing nonlinearity of current-voltage curves as extrapolated from
the insulating phase (circles and squares) and on one sample from
Ref.~\cite{li94} using a cutoff criterion,
$\sigma_{xx}^{-1}=10$~MOhm, at a temperature $\approx 25$~mK as
follows from the former method (diamonds). The solid lines are guides
to the eye. Digits indicate $\sigma_{xy}h/e^2$ for different
insulating phases. From Ref.~\cite{shashkin94b}.}
\end{figure}

The consequences of the method include (i)~as long as there occurs no
dramatic change in transport properties, this excludes the pinned
Wigner solid as the origin for the insulating phase at low electron
densities in available samples of low-disordered silicon MOSFETs;
(ii)~the metal-insulator phase diagram of Fig.~\ref{floating}(a) is
verified and substantiated; (iii)~the existence of a metal-insulator
transition in zero magnetic field is supported (see
section~\ref{true}); and (iv)~the bandwidth of the extended states in
the Landau levels is finite. All of these are also valid for
relatively low-disordered 2D carrier systems in GaAs/AlGaAs
heterostructures with a distinction that fractional quantum Hall
phases are involved. Yet, the topology of the phase diagram remains
unchanged including the oscillating behavior of the phase boundary
that restricts the low-density insulating phase (see
Fig.~\ref{gaas}). Additional confirmation of the percolation
transition to the low-density insulating phase in GaAs/AlGaAs
heterostructures was obtained by studies of the high-frequency
conductivity \cite{li94} and time-resolved photoluminescence of 2D
electrons \cite{kukushkin93}, as discussed in
Ref.~\cite{shashkin94b}.

\subsubsection{Finite bandwidth of extended states}

It was predicted two decades ago that the localization length
diverges as a power law at a single energy, $E^*$, which is the
center of the Landau level \cite{iordansky82,ando83,aoki83}:
$L(E)\propto |E-E^*|^{-s}$. An idea to check this prediction based on
low-temperature measurements of $\sigma_{xx}$ \cite{aoki85} was
quickly developed to a concept of single-parameter scaling
\cite{pruisken88}. It was suggested that the magnetoresistance tensor
components are functions of a single variable that is determined by
the ratio of the dephasing length, $L_d(T)\propto T^{-p/2}$ (where
$p$ is the inelastic-scattering-time exponent), and the localization
length. The concept was claimed to be confirmed by measurements of
temperature dependences of the peak width, $\Delta B$, in $\rho_{xx}$
(or $\sigma_{xx}$) and the maximum of $d\rho_{xy}/dB$ in a
highly-disordered 2D electron system in InGaAs/InP heterostructures,
which yielded $\Delta B\propto T^\kappa$, where $\kappa=p/2s\approx
0.4$ \cite{wei88}. Later, both deviations in the power law and
different exponents in the range between $\kappa=0.15$ and $\kappa=1$
were observed for other 2D carrier systems, different Landau levels,
and different disorder strengths
\cite{shahar95a,wang94,hughes94,wakabayashi89,koch91a,dolgopolov91a,koch91b,koch92,wei92,hwang93,engel93,wei94,wong95,shahar95b,pan97,shahar97,coleridge99,schaijk00,dunford00,dunford01,hohls01,hohls02a,hohls02b}.
Importantly, the scaling analysis of experimental data in question is
based on two unverified assumptions: (i)~zero bandwidth of the
extended states in the Landau levels; and (ii)~constancy of the
thermodynamic density of states in the scaling range. If either
assumption is not valid, this may lead at least to underestimating
the experimental value of exponent $\kappa$.

\begin{figure}
\scalebox{0.42}{\includegraphics[clip]{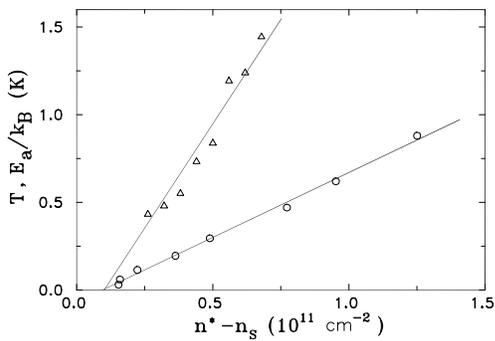}}
\caption{\label{width} Temperature dependence of the $\rho_{xx}$ peak
width ($n^*-n_s$) at half of the peak height counted from $n^*$
corresponding to $\nu^*=2.5$ (circles) and the behavior of the
activation energy (triangles) in a low-disordered silicon MOSFET in
$B=14$~T. From Ref.~\cite{shashkin94a}.}
\end{figure}

\begin{figure}
\scalebox{0.38}{\includegraphics[clip]{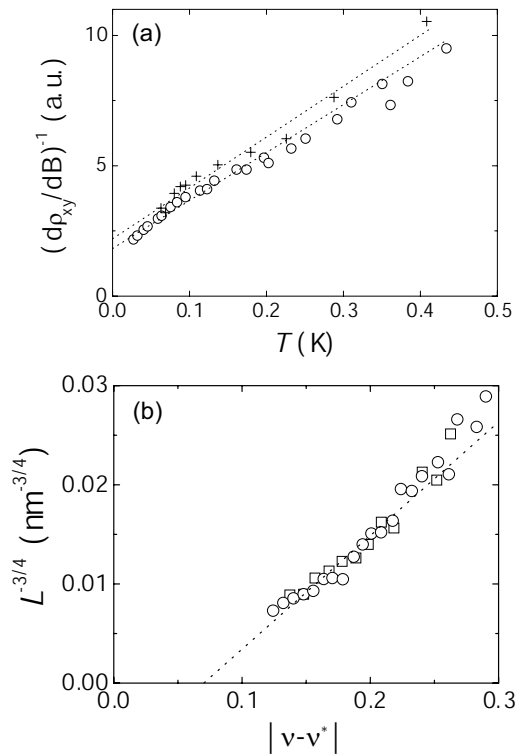}}
\caption{\label{wei} (a)~Temperature dependence of the $\rho_{xx}$
peak width as determined by the maximum $d\rho_{xy}/dB$ at
$\nu^*=1.5$ in a highly-disordered 2D electron system in GaAs/AlGaAs
heterostructures. Different symbols correspond to different runs. The
dashed lines are linear fits to the data. Adopted from
Ref.~\cite{wei92}. (b)~Localization length, determined by the
high-frequency conductivity in the variable-range-hopping regime in a
highly-disordered 2D electron system in GaAs/AlGaAs heterostructures,
as a function of the filling factor deviation from $\nu^*=2.5$
(squares) and $\nu^*=3.5$ (circles) towards the $\nu=3$ plateau. The
dashed line is a linear fit that is expected from classical
percolation approach with exponent $s=4/3$. Adopted from
Ref.~\cite{hohls01}.}
\end{figure}

The method of vanishing activation energy and vanishing nonlinearity
of current-voltage characteristics as extrapolated from the
insulating phase shows that the former assumption is not justified.
Moreover, measurements of the peak width in $\rho_{xx}$ as a function
of temperature in low-disordered silicon MOSFETs yield a linear
dependence which extrapolates as $T\rightarrow0$ to a finite peak
width, accordingly \cite{shashkin94a} (see Fig.~\ref{width}). Very
similar temperature (and frequency) dependences were observed in
highly-disordered 2D carrier systems in GaAs/AlGaAs heterostructures
\cite{balaban98,shahar98} and Ge/SiGe heterostructures
\cite{hilke97,arapov02}. It is noteworthy that a similar behavior is
revealed if the data from the publications, which claim the
observation of scaling, is plotted on a linear rather than
logarithmic scale (see, e.g., Fig.~\ref{wei}); finite values of the
peak width as $T\rightarrow0$ are even more conspicuous for the data
of Refs.~\cite{dolgopolov91a,koch91b,koch92}. The reason for the
ambiguity is quite simple: within experimental uncertainty, it is
difficult (on a logarithmic scale it is especially difficult) to
distinguish between sublinear/superlinear fits to the data and linear
fits which do not have to run to the origin.

Although lack of data in most of the above experimental papers does
not allow one to verify the validity of both assumptions, it is very
likely that there is no qualitative difference between all of the
discussed results. As a matter of fact, they can be described by a
linear (or weakly sublinear) temperature dependence with a finite
offset at $T=0$. Here comes an alternative and simple explanation of
the temperature dependence of the peak width in $\rho_{xx}$ in terms
of thermal broadening. Within percolation picture, if the activation
energy $E_a\sim k_BT$, the conduction is order of the maximum
$\sigma_{xx}$ so that the value of $\sim k_BT$ gives a thermal shift
of the effective mobility edge corresponding to the $\sigma_{xx}$
peak width \cite{shashkin94a}. Despite the concept of thermal
broadening has been basically ignored in the literature in search for
less trivial data interpretations, it looks as if no experimental
results go beyond this favoring the concept of single-parameter
scaling. Once the behavior of the localization length is not
reflected by the temperature-dependent peak width in $\rho_{xx}$, no
experimental support is provided for numeric calculations of the
localization length which give a somewhat larger exponent $s\approx
2$ compared to $s=4/3$ in classical percolation problem (see, e.g.,
Ref.~\cite{huckestein95}). The latter value of $s$ as well as the
behavior of the localization length in Fig.~\ref{Vc} have been
recently confirmed by measurements of the high-frequency conductivity
in the variable-range-hopping regime \cite{hohls01} (see
Fig.~\ref{wei}(b)).

Thus, the finding of finite bandwidth of the extended states in the
Landau levels \cite{shashkin94a,shashkin94b}, which is in obvious
contradiction to scaling arguments, is a firmly-established
experimental result. Curiously, it has got no theoretical
considerations for ten years.

\subsubsection{Hall insulator}

Deep in the insulating phases and at low temperatures the
variable-range-hopping regime occurs in which the conductivity
$\sigma_{xx}$ is small compared to its peak value
\cite{shklovskii84}. In this regime it was predicted that the
deviation, $\Delta\sigma_{xy}$, of $\sigma_{xy}$ from its quantized
value in strong magnetic fields is much smaller than
$\sigma_{xx}\propto\exp(-(T_0/T)^{1/2})$ \cite{wysokinski83}:
$\Delta\sigma_{xy}\propto\sigma_{xx}^\gamma$ with exponent
$\gamma\approx 1.5$; note that this is in contrast to a
straightforward linear relation in the activation regime, as inferred
from approximately the same behavior with temperature of the
$\rho_{xx}$ peak width and the maximum of $d\rho_{xy}/dB$. Later,
finite $\rho_{xy}$ contrasted by diverging $\rho_{xx}$ was found in
calculations of the $T=0$ magnetotransport coefficients in the
insulating phase with vanishing $\sigma_{xx}$ and $\sigma_{xy}$
\cite{viehweger90,viehweger91}. Such a behavior of $\rho_{xx}$ and
$\rho_{xy}$ indicates a special quadratic relation between
conductivities
\begin{equation}\sigma_{xy}\propto\sigma_{xx}^2.\label{special}\end{equation}
Moreover, it was shown that $\rho_{xy}$ is close to the classical
value ($B/n_sec$) \cite{zhang92} providing arguments for the
existence of a Hall insulator phase \cite{kivelson92}.

Values $\rho_{xy}$ close to $B/n_sec$ were experimentally found in
the low-density insulating phase, deviations from the classical Hall
line being attributed to possible admixture of $\rho_{xx}$
\cite{goldman88,pudalov93b,wakabayashi88,dorozhkin93,goldman93,sajoto93,kravchenko94b,pudalov94}.
Thus, the distinction of the Hall insulator phase from the quantum
Hall phases --- the absence of extended states below the Fermi level
--- becomes evident when expressed in terms of $\rho_{xx}$ and
$\rho_{xy}$.

It was empirically found in low-disordered silicon MOSFETs that the
lowest-filling-factor peak in $\sigma_{xx}$ plotted in the
($\sigma_{xy},\sigma_{xx}$) plane is close to a semicircle centered
at ($e^2/2h,0$) \cite{kravchenko91,dolgopolov92c,pudalov94}. The
semicircle law for the lowest-$\nu$ peak was reproduced in a
highly-disordered 2D hole system in a Ge/SiGe quantum well
\cite{hilke98,hilke99}. There, it was shown that the semicircle
relation originates directly from conductivity/resistivity tensor
inversion
\begin{equation}\sigma_{xx}^2+\left(\sigma_{xy}-\frac{e^2}{2h}\right)^2=\left(\frac{e^2}{2h}\right)^2+\frac{1-\rho_{xy}e^2/h}{\rho_{xx}^2+\rho_{xy}^2},\label{semi}\end{equation}
because the (narrow) $\sigma_{xx}$ peak in question is located at
filling factor just below $\nu=1$ (see, e.g., Fig.~\ref{floating})
where $\rho_{xy}$ is still close to $h/e^2$. Although this finding is
consistent with theories
\cite{levine83,khmelnitskii83,dykhne94,ruzin95,burgess00}, the
semicircle law does not seem universal if higher-$\nu$ peaks in
$\sigma_{xx}$ with different heights are involved
\cite{kravchenko91,dolgopolov91a,pudalov94,wei86}.

\subsection{Edge channel effects, direct measurements of the quantized Hall conductivity}
\label{edge}

\begin{figure}
\scalebox{0.4}{\includegraphics[clip]{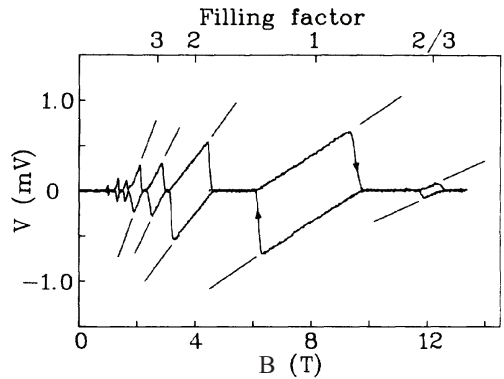}}
\caption{\label{charge} The induced voltage in a Corbino sample of a
GaAs/AlGaAs heterostructure in up- and down-sweeps of the magnetic
field. Also shown by straight lines are the expected slopes for
$\nu=2/3$, 1, 2, 3, and 4. From Ref.~\cite{dolgopolov92d}.}
\end{figure}

In a magnetically quantized 2D electron system, the Landau levels
bend up at the sample edges due to the confining potential, and edge
channels are formed where these intersect the Fermi energy (see,
e.g., Ref.~\cite{halperin82}). There rises a natural question whether
the current in the quantum Hall state flows in the bulk or at the
edges of the sample. Although the Hall conductivity $\sigma_{xy}$ was
not directly measured in early experiments on the quantum Hall
effect, it seemed obvious that this value corresponds to $\rho_{xy}$,
in agreement with the concept of currents that flow in the bulk
\cite{QHE87}; that stands to reason that finite $\sigma_{xy}$ would
give evidence for the existence of extended states in the Landau
levels \cite{levine83,halperin82}. This concept was challenged by the
edge current model \cite{buttiker88}. In the latter approach extended
states in the bulk are not crucial and the problem of current
distributions in the quantum Hall effect is reduced to a
one-dimensional task in terms of transmission and reflection
coefficients as defined by the backscattering current at the Fermi
level between the edges. Importantly, if the edge current contributes
significantly to the net current, conductivity/resistivity tensor
inversion is not justified, because the conductivities $\sigma_{xx}$
and $\sigma_{xy}$ are related to the bulk of the 2D electron system.

\begin{figure*}
\includegraphics[height=0.5\textheight]{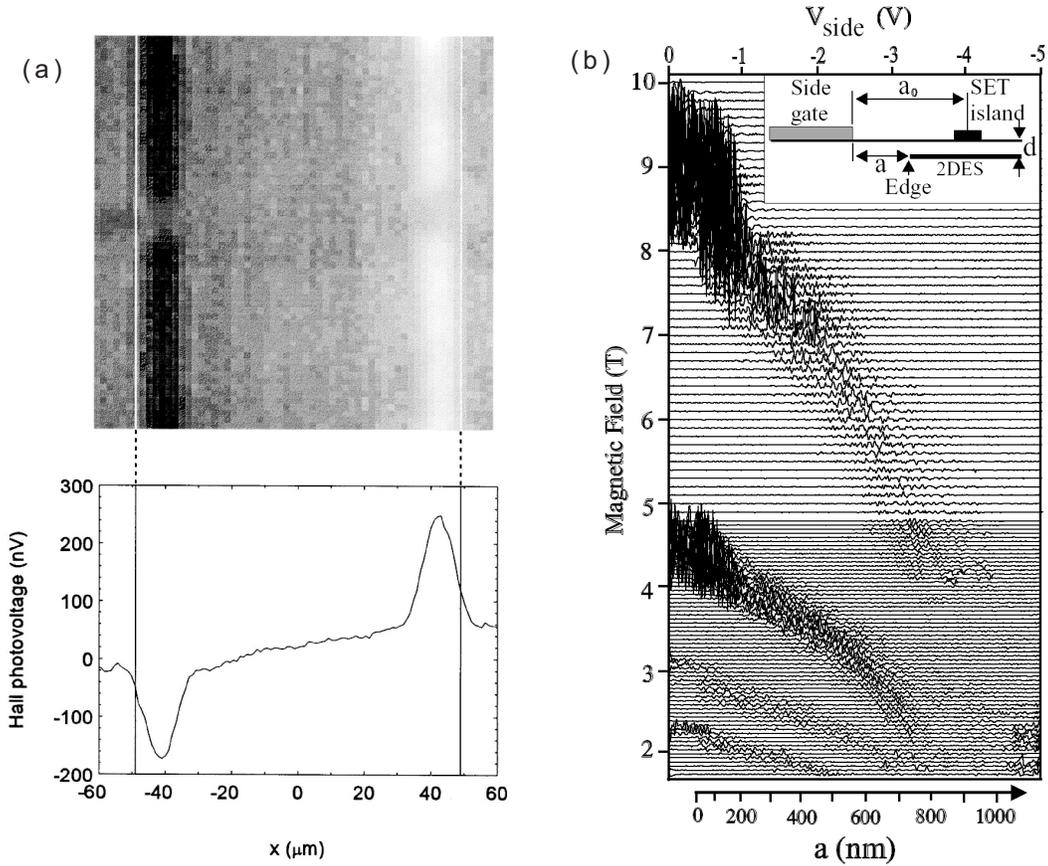}
\caption{\label{image} (a)~Image of the Hall photovoltage in a near
surface GaAs/AlGaAs heterojunction at $\nu=2$ (top) and a line scan
taken horizontally through the image (bottom). The spot size is
5~$\mu$m across. The physical edges of the sample are indicated by
vertical lines. From Ref.~\cite{shashkin97}. (b)~Traces of the
fluctuation part of the feedback signal vs the side-gate voltage for
shifting the edge in a GaAs/AlGaAs heterostructure at different
magnetic fields. Enhanced fluctuations in the feedback signal
indicate vanishing $\sigma_{xx}$, which corresponds to Hall current
channels (incompressible strips) for $\nu=1$, 2, 3, and 4. Adopted
from Ref.~\cite{wei98}.}
\end{figure*}

To verify whether or not the Hall conductivity is quantized, direct
measurements of $\sigma_{xy}$ were necessary excluding a possible
shunting effect of the edge currents. Being equivalent to Laughlin's
gedanken experiment \cite{laughlin81,widom82}, such measurements were
realized using the Corbino geometry which allows separation of the
bulk contribution to the net current
\cite{dolgopolov90,dolgopolov91b,dolgopolov92d,dolgopolov93,jeanneret95,watts98,honold99,dolgopolov01}.
A Hall charge transfer below the Fermi level between the coasts of a
Corbino sample is induced by magnetic field sweep through the
generated azimuthal electric field. If $\sigma_{xx}\rightarrow0$, no
discharge occurs allowing determination of the transferred charge,
\begin{equation}Q=\sigma_{xy}\pi r_{\text{eff}}^2c^{-1}\delta B,\label{Q}\end{equation}
where $r_{\text{eff}}$ is the effective radius. The induced voltage,
$V=Q/C$, which is restricted due to a large shunting capacitance,
$C$, changes linearly with magnetic field with a slope determined by
$\sigma_{xy}$ in the quantum Hall states until the dissipationless
quantum Hall state breaks down (see Fig.~\ref{charge}). The fact that
the quantization accuracy of $\sigma_{xy}$ (about 1\%) is worse
compared to $\rho_{xy}$ may be attributed to non-constancy of the
effective area in not very homogeneous samples. Thus, the Hall
current in the quantum Hall effect flows not only at the edges but
also in the bulk of the 2D electron system through the extended
states in the filled Landau levels.

Apparently, the dissipative backscattering current in Hall bar
samples should be balanced by Hall current in the filled Landau
levels resulting in a longitudinal potential drop
\cite{dolgopolov91c}. This point makes significant contribution to
the edge current model.

From an experimental viewpoint, all edge channel effects proceed from
slow equilibration (at macroscopic distances) between the
electrochemical potentials of different edge states including the
state in the bulk. As long as such an equilibration occurs at the
edges at the Fermi level, the applicability of the edge state model
is justified. The approach accounts for effects observed in
conventional transport experiments, which include the nonlocal
resistance and the effects related to contacts/reservoirs (see, e.g.,
Ref.~\cite{haug93}). However, particular potential profiles at the
edge and current distributions can be probed only using
non-destructive spatially-resolved imaging techniques
\cite{fontein91,kent92,merz93,shashkin94c,shashkin94d,haren95a,haren95b,shashkin97,wei98,tessmer98,mccormick99,yacoby99,shashkin99,zhitenev00,finkelstein00a,finkelstein00b,woodside01,zalinge01,glicofridis02};
note that many of the so-revealed inhomogeneous samples show quite
good magnetotransport characteristics. Contrary to standard
considerations of the edge channels in terms of skipping orbits for a
confining potential that is sharp on the magnetic length, $l_B=(\hbar
c/eB)^{1/2}$, it turns out that in most samples the potential profile
at the edge is smooth and spans over much larger distances than
$l_B$. Edge regions corresponding to the scale of confining potential
$\approx 10$~$\mu$m were visualized in Hall photovoltage optical
imaging experiments on standard Hall bar samples \cite{shashkin97}
(see Fig.~\ref{image}(a)). Since the Hall electric field is nearly
constant, even if some field enhancement occurs near the edges
\cite{fontein91}, the edge current contribution can be appreciable
depending on a particular sample.

For a soft confining potential, edge channels are also referred to as
compressible and incompressible strips whose spacing is determined by
the electron density gradient \cite{chklovskii92}. This is very
similar to the long-standing phenomenon of Hall current pinch: given
electron density gradients, the Hall current basically flows in
narrow channels (or incompressible strips) determined by the minimum
$\sigma_{xx}$, their position in the sample being controlled, e.g.,
by magnetic field
\cite{haren95a,haren95b,ebert85,pudalov85,shashkin86,wiegers99}. When
located at the edge, the pinch of Hall current becomes identical with
the subject of Ref.~\cite{chklovskii92}. The Hall current channels at
the edge were imaged using a single-electron transistor as a probe
for the local $\sigma_{xx}$ \cite{wei98} (see Fig.~\ref{image}(b)).
Applying a negative voltage to a side gate leads to a shift of the
edge of the 2D electron system towards the probe, thereby creating a
line scan of the local $\sigma_{xx}$ across the sample edge.
Vanishing $\sigma_{xx}$ is indicated by enhanced fluctuations in the
feedback signal, the feedback circuit being used to keep the current
through the single-electron transistor constant by controlling its
voltage relative to the 2D electron system.

With respect to the preceding subsections, insignificance of edge
channel effects in transport experiments is verified in a usual way
by coincidence of the results obtained in Hall bar and Corbino
geometries.

\subsection{True zero-field metal-insulator transition, phase boundary in parallel magnetic fields}
\label{true}

As has been discussed above, the existence of extended states in
quantizing magnetic fields is established by two independent
experimental methods: (i)~quantization of $\sigma_{xy}$; and
(ii)~vanishing activation energy and vanishing nonlinearity of
current-voltage characteristics as extrapolated from the insulating
phase. Theory is generally in agreement with this even though there
are unresolved problems with finite bandwidth of the extended states
in the Landau levels. In contrast, no extended states are expected in
zero magnetic field, at least, for weakly-interacting 2D electron
systems. The second experimental criterion, however, results in an
opposite conclusion although it does not have absolute credibility
alone. To sort it out, further support by independent experimental
verifications is needed.

\begin{figure}
\scalebox{0.64}{\includegraphics[clip]{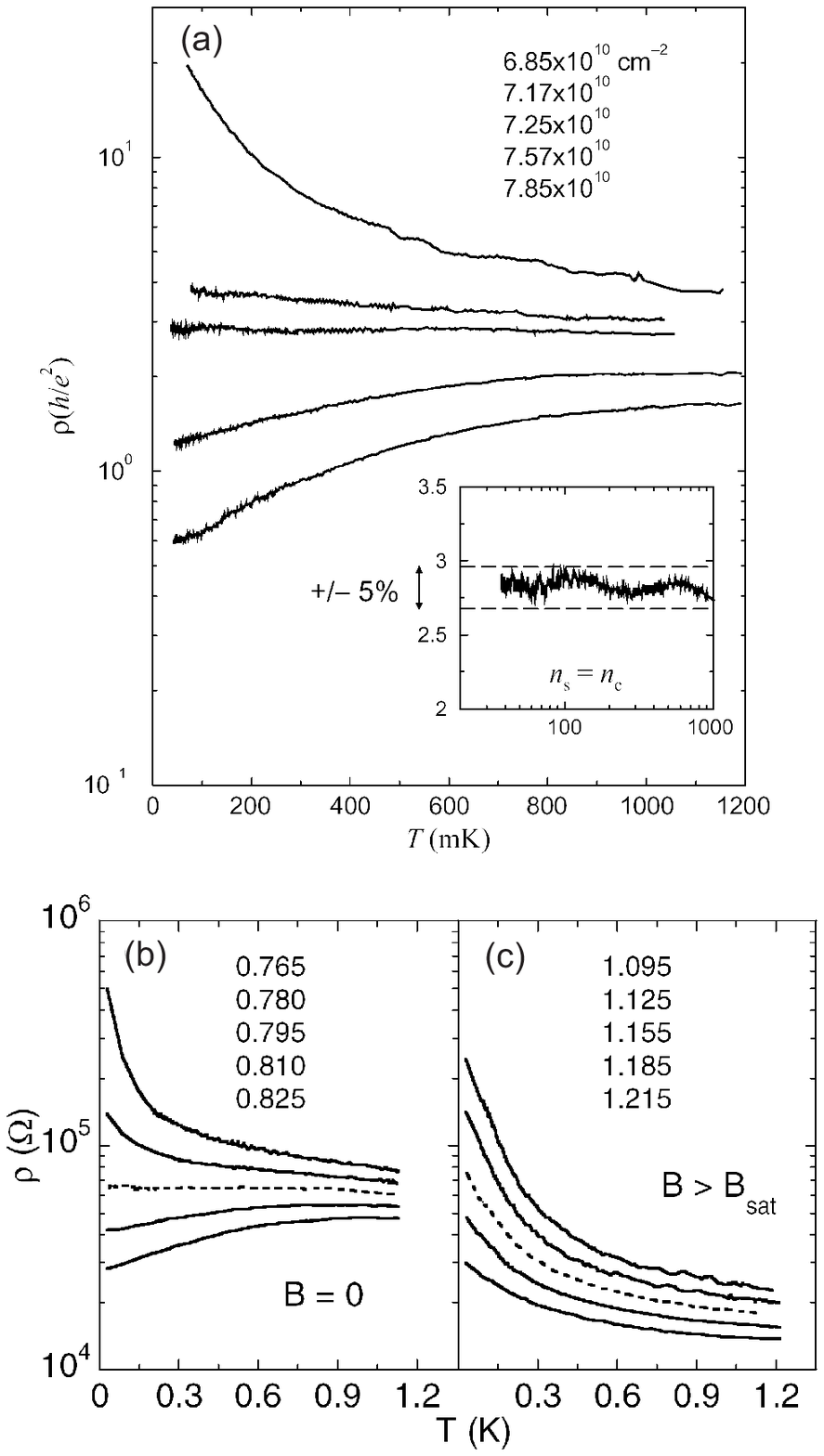}}
\caption{\label{flat} (a)~Resistivity as a function of temperature at
different electron densities in a low-disordered silicon MOSFET. The
inset shows the middle curve on an expanded scale. Adopted from
Ref.~\cite{kravchenko00b}. (b, c)~Temperature dependence of the
resistivity of a low-disordered silicon MOSFET at different electron
densities near the metal-insulator transition in zero magnetic field
(b) and in a parallel magnetic field of 4~T (c). The electron
densities are indicated in units of $10^{11}$~cm$^{-2}$. From
Ref.~\cite{shashkin01b}.}
\end{figure}

\begin{figure}
\scalebox{0.55}{\includegraphics[clip]{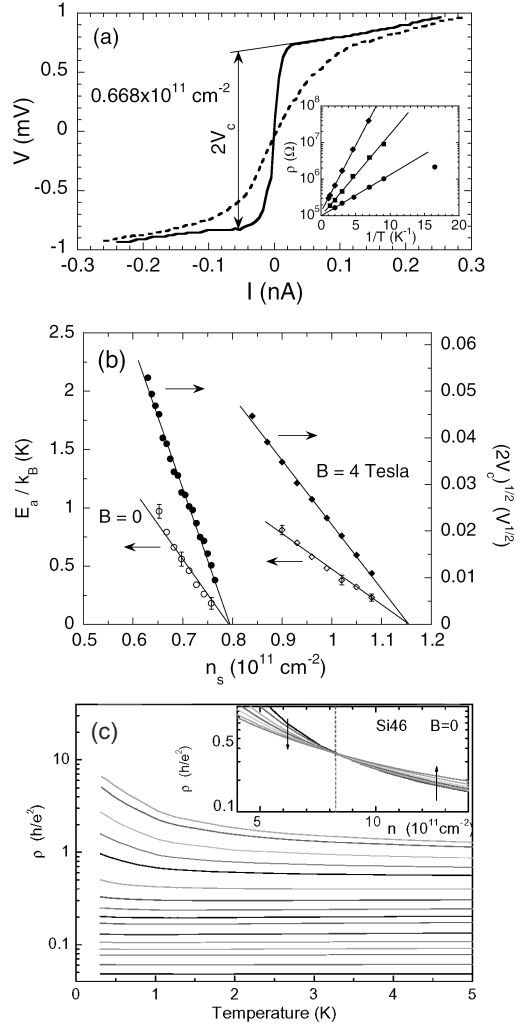}}
\caption{\label{nc2} (a)~Current-voltage characteristics in zero
magnetic field at $\approx 30$ (solid line) and 211~mK (dashed line)
for the same silicon MOSFET as in Fig.~\ref{flat}(b, c); note that
the threshold voltage is practically independent of temperature. An
Arrhenius plot of the resistivity in the insulating phase is
displayed in the inset for different values of $B_\parallel$ and
$n_s$. From Ref.~\cite{shashkin01b}. (b)~Activation energy and square
root of the threshold voltage as a function of electron density in
zero magnetic field (circles) and in a parallel magnetic field of 4~T
(diamonds). The critical densities correspond to the dashed lines in
Fig.~\ref{flat}(b, c). From Ref.~\cite{shashkin01b}. (c)~Resistivity
versus temperature in a strongly-disordered silicon MOSFET at the
following electron densities: 3.85, 4.13, 4.83, 5.53, 6.23, 7.63,
9.03, 10.4, 11.8, 13.2, 16.0, 18.8, 21.6, 24.4, 30.0, and $37.0\times
10^{11}$~cm$^{-2}$. The $\rho(n_s)$ isotherms are shown in the inset.
Adopted from Ref.~\cite{pudalov01}.}
\end{figure}

Another criterion is based on analysis of the temperature dependences
of the resistivity in $B=0$. Provided these are strong, those with
positive (negative) derivative $d\rho/dT$ are indicative of a metal
(insulator)
\cite{kravchenko94a,kravchenko95a,kravchenko96,popovic97,coleridge97};
note that in the vicinity of the transition, $\rho(T)$ dependences
obey the scaling law with exponent $\kappa\approx1$, which is
consistent with the concept of thermal broadening/shift by the value
$\sim k_BT$ of the effective mobility edge in the insulating phase
(see section~\ref{similarity}). If extrapolation of $\rho(T)$ to
$T=0$ is valid, the critical point for the metal-insulator transition
is given by $d\rho/dT=0$. In a low-disordered 2D electron system in
silicon MOSFETs, the resistivity at a certain electron density shows
virtually no temperature dependence over a wide range of temperatures
\cite{sarachik99,kravchenko00b,jaroszynski02} (see
Fig.~\ref{flat}(a)). This curve separates those with positive and
negative $d\rho/dT$ nearly symmetrically at temperatures above 0.2~K
\cite{simonian97a}. Assuming that it remains flat down to $T=0$, one
obtains the critical point $n_c$ which corresponds to a resistivity
$\rho\approx 3h/e^2$ \cite{abrahams01}.

Recently, these two criteria have been applied simultaneously to the
2D metal-insulator transition in low-disordered silicon MOSFETs
\cite{shashkin01b,jaroszynski02}. In zero magnetic field, both
methods yield the same critical density (see Figs.~\ref{flat}(b) and
\ref{nc2}(b)). Since one of the methods is temperature independent,
this equivalence strongly supports the existence of a true
metal-insulator transition in $B=0$. This also adds confidence that
the curve with zero derivative $d\rho/dT$ will remain flat (or at
least will retain finite resistivity value) down to zero temperature.
Additional confirmation in favor of true zero-field metal-insulator
transition is provided by magnetic measurements as described in the
next section.

In the presence of a parallel magnetic field, $B_\parallel$, the
outcome is very different. With increasing parallel field the
transition point, $n_c(B_\parallel)$, determined from the vanishing
nonlinearity and activation energy, shifts approximately linearly to
higher electron densities, saturating above a critical field,
$B_c\approx 3$~T, at a constant value which is approximately 1.5
times higher than that in zero field. Note that a similar suppression
of the metallic state was observed using a cutoff criterion,
$\rho=100$~kOhm \cite{dolgopolov92c}. In the metallic phase the
saturation of the resistance with parallel field signals the onset of
full spin polarization of the 2D electrons, as inferred from an
analysis of the Shubnikov-de~Haas oscillations in tilted magnetic
fields \cite{okamoto99,vitkalov00,vitkalov01b}. One expects that the
2D electron system is spin polarized at parallel fields
$B_\parallel>B_c$, and that the observed phase boundary shift is a
spin effect. At the so-determined critical density
$n_c(B_\parallel)$, the exponential divergence of the resistivity as
$T\rightarrow0$ ends, although $d\rho/dT$ remains negative at least
for $B_\parallel>B_c$ (see Figs.~\ref{flat}(c) and \ref{nc2}(b)). In
sharp contrast with the zero-field case, not only are the $\rho(T)$
curves in the field nonsymmetric about the middle curve in
Fig.~\ref{flat}(c), but all of them have negative derivatives
$d\rho/dT$ in the entire temperature range, although the values of
$\rho$ are comparable to those in the $B=0$ case. The metallic
($d\rho/dT>0$) temperature dependence of the resistivity, observed at
higher electron densities in parallel magnetic fields, is weak so
that the derivative method does not yield a critical density at
$B_\parallel>B_c$. Its failure leaves uncertain the existence of a
true metal-insulator transition in a parallel magnetic field.

\begin{figure*}
\includegraphics[height=0.5\textheight]{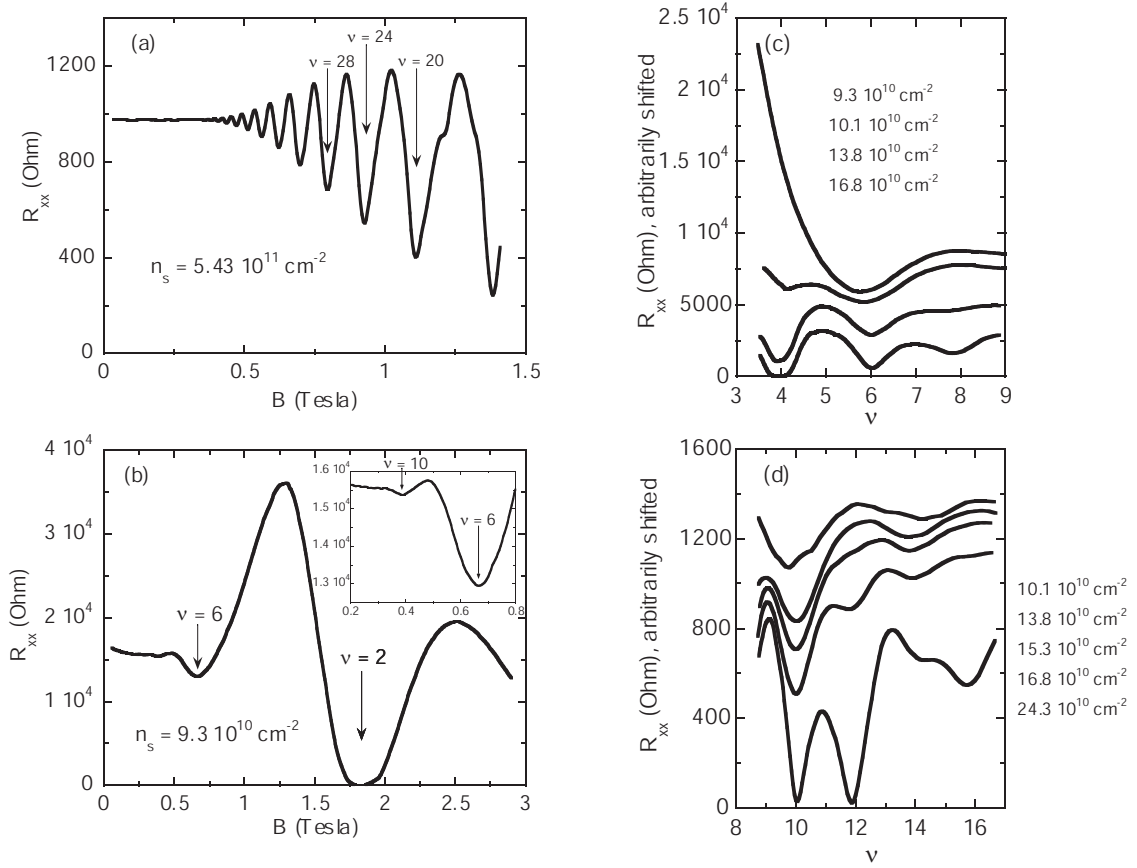}
\caption{\label{SdH1} Shubnikov-de~Haas oscillations in a
low-disordered silicon MOSFET at a temperature $\approx40$~mK for
high (a) and low (b) electron densities. (c, d) Evolution of the
Shubnikov-de~Haas oscillations with electron density in two ranges of
filling factors. The curves are arbitrarily shifted along the
$y$-axis for clarity. From Ref.~\cite{kravchenko00a}.}
\end{figure*}

A very similar conclusion holds for 2D electron systems with higher
disorder in zero magnetic field (see section~\ref{floating-up}). In
this case the metallic ($d\rho/dT>0$) behavior is also suppressed
\cite{papadakis98,hanein98,simmons98,mills99,yoon99,simmons00,noh03,pudalov01}
or disappears entirely, and extrapolation of the weak $\rho(T)$
dependences to $T=0$ is not justified invalidating the derivative
criterion for the critical point for the metal-insulator transition
(see Fig.~\ref{nc2}(c)). This is noteworthy that owing to its
simplicity, the derivative method is widely used for describing
metallic ($d\rho/dT>0$) and insulating ($d\rho/dT<0$) temperature
dependences of resistance in a restricted temperature range. To avoid
confusion with metallic and insulating phases, however, one should
employ alternative methods for determining the metal-insulator
transition point. Such methods including a vanishing activation
energy and noise measurements have been applied to highly-disordered
2D carrier systems \cite{jaroszynski02,bogdanovich02,leturcq03}.
Being similar, they yield lower critical densities for the
metal-insulator transition compared to those obtained using formally
the derivative criterion. This simply reflects the fact that the
metallic ($d\rho/dT>0$) behavior is suppressed, the critical density
$n_c$ increasing naturally with disorder strength (see
Fig.~\ref{diagram}).

\section{MANY-BODY PHENOMENA IN DILUTE 2D ELECTRON SYSTEMS}

The resistivity drop with decreasing temperature in a low-disordered
2D electron system in silicon MOSFETs in zero magnetic field being
strong compared to the metallic $\rho(T)$ expected from the
temperature-dependent screening theories
\cite{stern80,gold86,dassarma86,dassarma99}, it was attributed to a
manifestation of strong electron-electron interactions
\cite{kravchenko94a}. Recently, the underlying physics of the effect
has been clarified. At low electron densities, a strongly enhanced
ratio $gm$ of the spin and the cyclotron splittings has been found
indicating that the 2D system behaves well beyond the weakly
interacting Fermi liquid \cite{kravchenko00a}. Experimental results
have also shown that it is the effective mass that increases sharply
at low electron densities and is related to the anomalous rise of the
resistivity with temperature \cite{shashkin02b}. In view of $T=0$
quantum phase transitions, while for the insulating phase a
transition is signaled by the diverging localization length, the
interaction-enhanced mass may be a similar indicator for the metallic
phase.

\subsection{Strong increase of the product $gm$ near the metal-insulator transition, possible ferromagnetic transition}
\label{strong}

\subsubsection{Beating pattern of Shubnikov-de~Haas oscillations}

Electron-electron interactions give rise to a renormalization of the
Fermi-liquid parameters including the effective mass and $g$ factor
\cite{landau57}. Tracing Shubnikov-de~Haas oscillation minima in a 2D
electron system in tilted magnetic fields, it is easy to determine
the ratio $gm$ of spin and cyclotron splittings which is proportional
to the spin susceptibility $\chi$. In the range of high electron
densities $\ge2\times 10^{11}$~cm$^{-2}$ in silicon MOSFETs, moderate
enhancements of $gm$ by a factor of $\le2.5$ were observed
\cite{okamoto99,fang68,smith72}, which is consistent with the concept
of weakly interacting Fermi liquid.

At low electron densities in low-disordered silicon MOSFETs in
perpendicular magnetic fields, the Shubnikov-de~Haas oscillation
minima corresponding to the cyclotron splittings ($\nu=4$, 8, 12,
16,...) were found to disappear as the electron density is reduced
\cite{kravchenko00a} (see Fig.~\ref{SdH1}). Disregarding the minimum
for the valley splitting at $\nu=1$, only minima corresponding to the
spin splittings ($\nu=2$, 6, 10, 14,...) remain close to the
metal-insulator transition which occurs in the studied samples at
$n_c\approx 8\times 10^{10}$~cm$^{-2}$. These results show that as
one approaches the metal-insulator transition, the cyclotron gaps
(which are equal to the difference between the cyclotron and spin
splittings, ignoring the valley splitting) become smaller than the
spin gaps and eventually vanish. The condition for vanishing is
coincidence of the spin and the cyclotron splittings, or $gm/2m_e=1$
(where $m_e$ is the free electron mass), which is higher by more than
a factor of 5 than the value of this ratio in bulk silicon,
$gm/2m_e=0.19$. The effect cannot be explained in the framework of
the many-body enhancement of spin gaps in a perpendicular magnetic
field \cite{ando74,bychkov81,kallin84,macdonald86,smith92} because
the disappearance of the cyclotron gaps in a wide range of magnetic
fields would require an enhanced $g$ factor which is independent of
magnetic field. This implies that the product $gm$ is nearly
field-independent and approximately equal to its many-body enhanced
zero-field value (see section~\ref{determining}). Thus, the spin
susceptibility $\chi\propto gm$ is strongly enhanced near the
metal-insulator transition.

Similar experiments in tilted magnetic fields cannot provide high
accuracy in determining the behavior of the renormalized $gm$ at low
electron densities because there are too few Shubnikov-de~Haas
oscillations near the metal-insulator transition. High accuracy was
attained in experiments on the parallel-field magnetotransport.

\subsubsection{Scaling the parallel-field magnetoresistance and other methods}

\begin{figure}
\scalebox{0.64}{\includegraphics[clip]{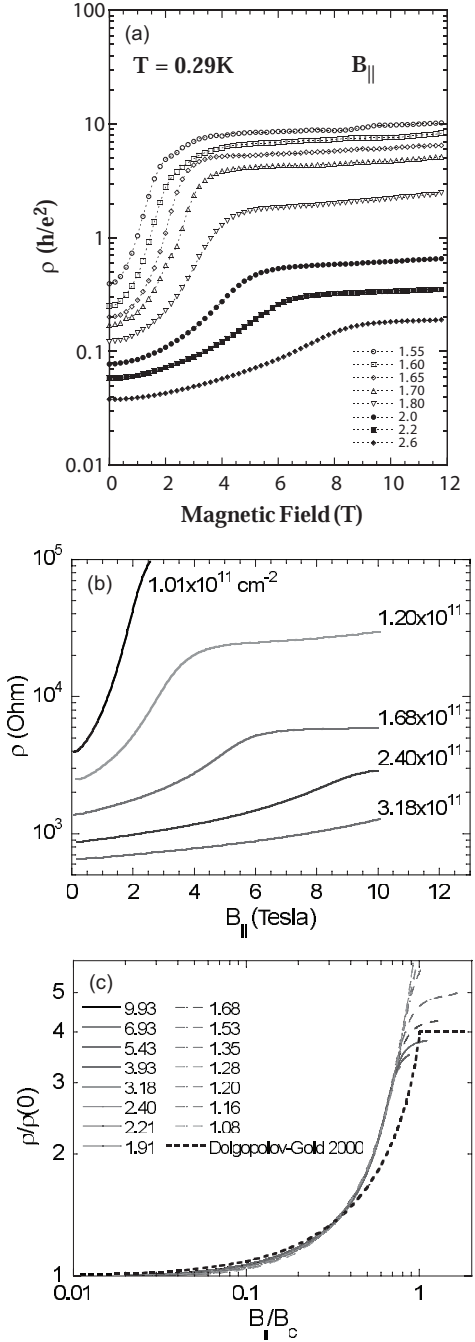}}
\caption{\label{saturation} (a)~Resistivity versus parallel magnetic
field measured on a low-disordered silicon MOSFET. Different symbols
correspond to gate voltages from 1.55 to 2.6~V, or to densities from
1.01 to $2.17\times 10^{11}$~cm$^{-2}$. Adopted from
Ref.~\cite{pudalov97}. (b)~Low-temperature magnetoresistance of a
low-disordered 2D electron system in silicon MOSFETs in parallel
magnetic fields at different densities above $n_c$ for the $B=0$
metal-insulator transition. From Ref.~\cite{shashkin01a}. (c)~Scaled
curves of the normalized magnetoresistance vs $B_\parallel/B_c$. The
electron densities are indicated in units of $10^{11}$~cm$^{-2}$.
Also shown by a dashed line is the normalized magnetoresistance
calculated by Dolgopolov and Gold \cite{dolgopolov00}. From
Ref.~\cite{shashkin01a}.}
\end{figure}

As the thickness of the 2D electron system in silicon MOSFETs is
small compared to the magnetic length in accessible fields, the
parallel magnetic field couples largely to the electrons' spins while
the orbital effects are suppressed. The resistance in dilute silicon
MOSFETs was found to be isotropic with respect to in-plane magnetic
field and rise steeply with the field saturating to a constant value
above a critical field $B_c$ which depends on electron density
\cite{simonian97b,pudalov97,pudalov02a} (see
Fig.~\ref{saturation}(a)). As has been mentioned in
section~\ref{true}, the saturation field $B_c$ corresponds to the
onset of full spin polarization of the electron system
\cite{okamoto99,vitkalov00,vitkalov01b}.

\begin{figure*}
\includegraphics[height=0.5\textheight]{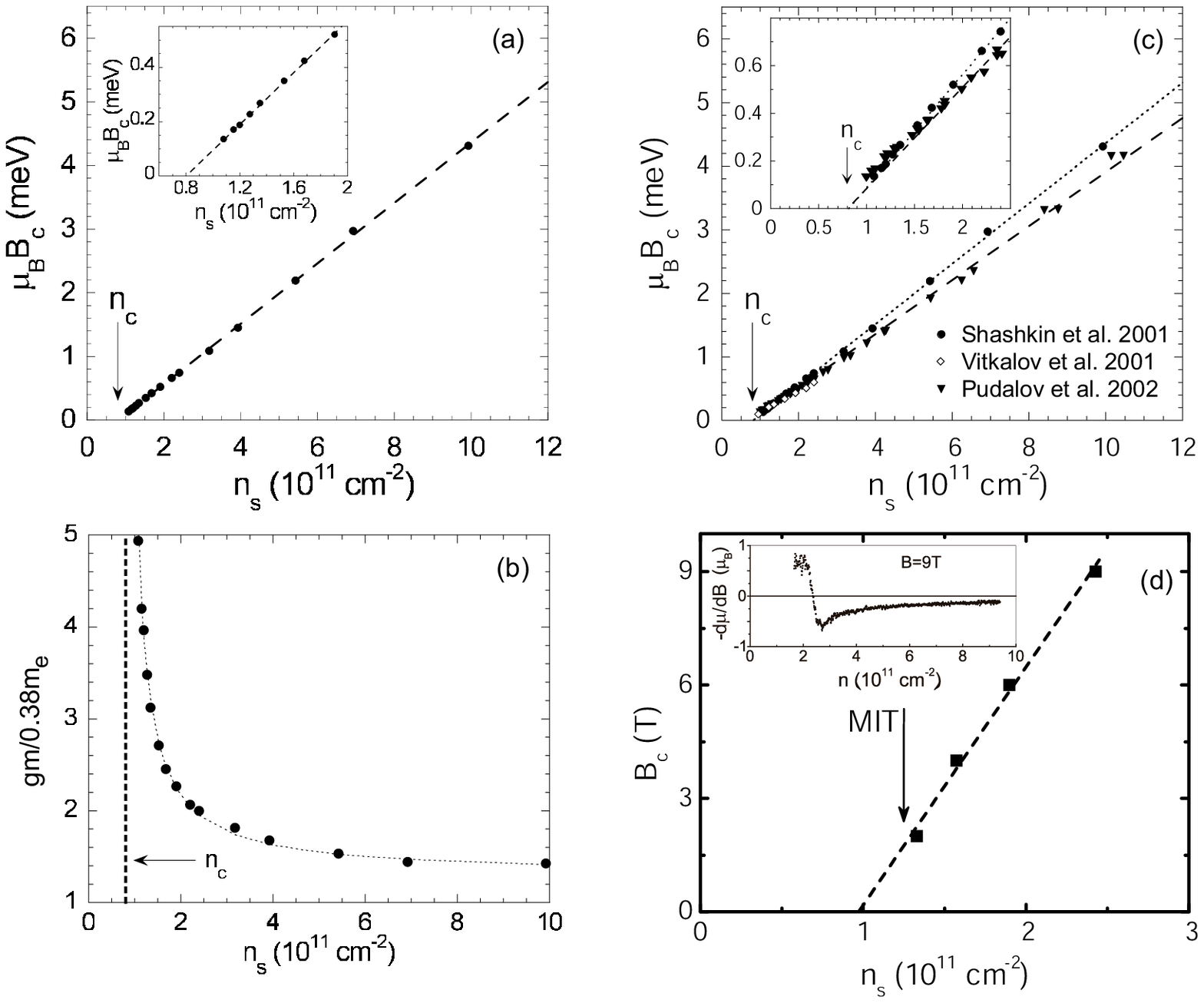}
\caption{\label{Bc} (a)~Dependence of the field $B_c$ on electron
density in a low-disordered silicon MOSFET. The dashed line is a
linear fit. From Ref.~\cite{shashkin01a}. (b)~The product $gm$ versus
electron density obtained from the data for $B_c$. From
Ref.~\cite{shashkin01a}. (c)~Polarization field as a function of
electron density obtained by different groups
\cite{shashkin01a,vitkalov01a,pudalov02b}. The critical density $n_c$
in the samples of Ref.~\cite{shashkin01a} is indicated. From
Ref.~\cite{shashkin02a}. (d)~Polarization field vs electron density
in a highly-disordered silicon MOSFET as determined from
$d\mu/dB_\parallel=0$ at 0.2~K. The dashed line is a linear fit. The
evaluated position of the metal-insulator transition is indicated.
The derivative $d\mu/dB_\parallel$ as a function of electron density
in a parallel magnetic field of 9~T at a temperature of 100~mK is
displayed in the inset. Adopted from Ref.~\cite{prus03}.}
\end{figure*}

In a low-disordered 2D electron system in silicon MOSFETs, it was
found that the normalized magnetoresistance, measured at different
electron densities in the low-temperature limit in which
$\rho(B_\parallel)$ becomes temperature independent, collapses onto a
single curve when plotted as a function of $B_\parallel/B_c$, where
the scaling parameter $B_c$ is normalized to correspond to the
saturation/polarization field \cite{shashkin01a} (see
Fig.~\ref{saturation}(b, c)). The scaling breaks down when one
approaches the metal-insulator transition where the magnetoresistance
depends strongly on temperature even at the lowest experimentally
achievable temperatures. Note that the observed scaling dependence is
described reasonably well by the theoretical dependence of
$\rho/\rho(0)$ on the degree of spin polarization,
$\xi=gm\mu_BB_\parallel/\pi\hbar^2 n_s=B_\parallel/B_c$, based on the
spin-polarization-dependent screening of a random potential
\cite{dolgopolov00}. The field $B_c$ is proportional, with high
precision, to the deviation of the electron density from its critical
value: $B_c\propto(n_s-n_c)$ (see Fig.~\ref{Bc}(a)). The procedure
used provides high accuracy for determining the functional form of
$B_c(n_s)$, even though the absolute value of $B_c$ is determined not
so accurately. Since the strong increase of the product $gm$ at low
electron densities (see Fig.~\ref{Bc}(b)), which follows from the
$B_c(n_s)$ dependence, is in agreement with the enhanced $gm$
obtained by Shubnikov-de~Haas oscillations, the band tail of
localized electron states is small and the clean limit occurs.
Therefore, the tendency for $B_c$ to vanish at a finite electron
density $n_\chi$ close to $n_c$ gives evidence in favor of the
existence of a ferromagnetic transition in this electron system
indicating that the metal-insulator transition is driven by
interactions \cite{shashkin01a}. It signifies that on the phase
diagram of Fig.~\ref{diagram}, the vicinity of either tricritical
point is reached (see section~\ref{wigner}).

A similar conclusion about possible spontaneous spin polarization was
drawn based on a scaling of magnetoconductivity data in similar
samples at different electron densities and temperatures
\cite{vitkalov01a}. Results for the strongly enhanced $gm$ were
corroborated in detailed studies of Shubnikov-de~Haas oscillations in
dilute silicon MOSFETs with higher disorder in tilted magnetic fields
\cite{pudalov02b} (see Fig.~\ref{Bc}(c)). The agreement between all
three sets of data is remarkable, especially if one takes into
account that different groups used different methods, different
samples, and different field/spin-polarization ranges
\cite{shashkin02a}. This also indicates that the electron density
$n_\chi\approx8\times 10^{10}$~cm$^{-2}$ is sample independent, in
contrast to the critical density $n_c$ for the metal-insulator
transition. Obviously, for the spin susceptibility $\chi\propto gm$
to diverge at $n_s=n_\chi$, the extrapolation of $B_c(n_s)$ to zero
must be valid. To verify its validity, accurate data at lower
densities, lower temperatures, and on much less disordered samples
are needed \cite{shashkin02a,vitkalov02,sarachik03}.

Thermodynamic investigations of the spin susceptibility, based on
measurements of the chemical potential change with parallel magnetic
field, $d\mu/dB_\parallel$, were performed in highly-disordered
silicon MOSFETs, as inferred from the considerably higher densities
for the metal-insulator transition \cite{prus03} (see
Fig.~\ref{Bc}(d), cf. Fig.~\ref{Bc}(c)). As compared to the clean
regime, the obtained dependence of the polarization field $B_c$ on
$n_s$ in Fig.~\ref{Bc}(d) is shifted to appreciably higher electron
densities caused by local moments in the band tail
\cite{pudalov02a,prus03,dolgopolov02a,gold02}. The band tail effects
thus become crucial in parallel-field experiments on
highly-disordered 2D electron systems.

\subsubsection{Other 2D carrier systems}

\begin{figure}
\scalebox{0.42}{\includegraphics[clip]{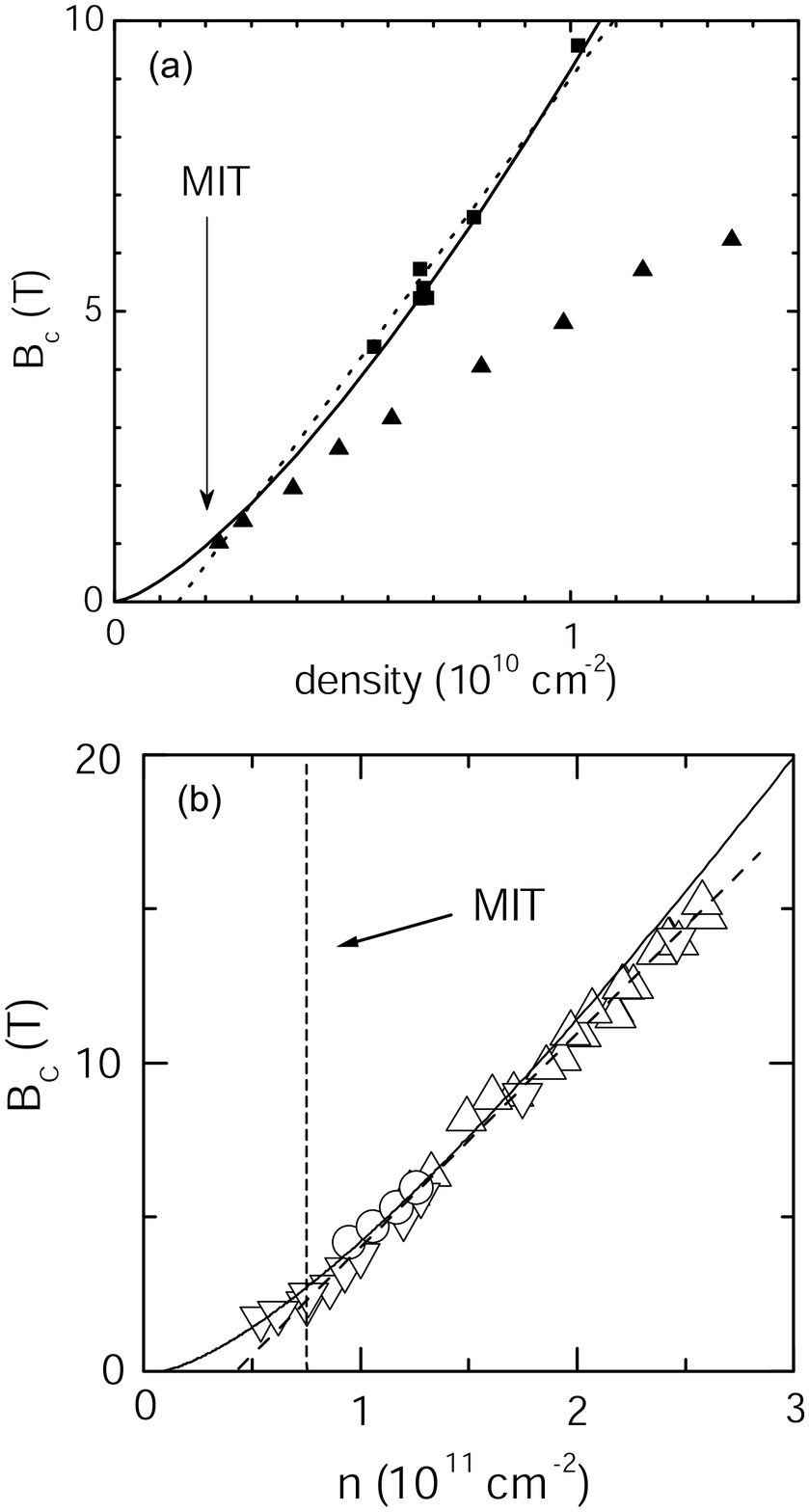}}
\caption{\label{zhu} (a)~Data for $B_c$ as a function of electron
density in a dilute GaAs/AlGaAs heterostructure obtained by
Shubnikov-de~Haas oscillations (squares) and parallel-field
magnetoresistance (triangles). The power-law fit $B_c\propto
n_s^{1.4}$ (solid line) is compared to the linear fit (dashed line).
Also shown is the evaluated position of the metal-insulator
transition. Adopted from Ref.~\cite{zhu03}. (b)~Dependence of $B_c$
on electron density obtained by measurements of the parallel-field
magnetoresistance in highly-disordered samples of narrow AlAs quantum
wells. The solid curve represents the quantum Monte Carlo calculation
for a disorder-free 2D electron system \cite{attaccalite02}. The
dashed line is a linear fit. The evaluated position of the
metal-insulator transition is indicated. Adopted from
Ref.~\cite{vakili04}.}
\end{figure}

In dilute GaAs/AlGaAs heterostructures, a similar enhancement of the
spin susceptibility at low electron densities was found by an
analysis of the Shubnikov-de~Haas oscillations \cite{zhu03} (see
Fig.~\ref{zhu}(a)). The thickness of the 2D carrier system in
GaAs/AlGaAs heterostructures is relatively large, which leads to an
increase of the effective mass with parallel magnetic field
\cite{zhu03,batke86,sarma00,tutuc02,noh02,tutuc03}. As a result, the
value of polarization field obtained from the parallel-field
magnetoresistance becomes strongly reduced as the electron density
increases. Disregarding this reduction, both data sets determine the
$B_c(n_s)$ dependence whose critical behavior is not so evident,
possibly because the lowest experimentally reached densities are
still too high. Due to the lower effective mass, higher dielectric
constant, and the absence of valley degeneracy, the same interaction
strength $r_s^*=E_{ee}/E_F$ in the 2D electron system in GaAs/AlGaAs
heterostructures is expected to be achieved at densities about two
orders of magnitude lower than in silicon MOSFETs. Therefore, the
critical region is expected at densities $n_s<10^9$~cm$^{-2}$ which
have not yet been accessed in currently available samples of
GaAs/AlGaAs heterostructures.

The orbital effects in parallel magnetic fields can be avoided by
using narrow quantum wells. A 2D electron system in narrow AlAs
quantum wells is similar to that in silicon MOSFETs, except that in
the former the valley degeneracy is absent \cite{vakili04}. The
critical region expected at densities $n_s<2\times 10^{10}$~cm$^{-2}$
is exceeded strongly by the lowest accessible electron densities in
narrow AlAs quantum wells with high disorder (see Fig.~\ref{zhu}(b)).
Note that the data points in the insulating phase reflect the physics
of local moments in the band tail
\cite{pudalov02a,prus03,dolgopolov02a,gold02}, which is different
from that of the metallic phase.

Being very similar to silicon MOSFETs, a 2D electron system in
Si/SiGe quantum wells is different by the higher dielectric constant
and the presence of a spacer. Moreover, it is distinguished from
other systems by its remote-doping scattering as indicated by the
small parallel-field magnetoresistance \cite{dolgopolov04}. A similar
increase of the spin susceptibility at low densities was observed in
this electron system, the lowest achievable densities also being well
above the expected critical region at $n_s<4\times 10^{10}$~cm$^{-2}$
\cite{okamoto04}. So, in all the studied dilute 2D electron systems
other than silicon MOSFETs, too high disorder and, hence, too high
densities for the metal-insulator transition (see Fig.~\ref{diagram})
mask possible critical behavior of the spin susceptibility.

\subsection{Determining separately the effective mass and $g$ factor}
\label{determining}

\subsubsection{Slope of the metallic temperature dependence of conductivity in zero magnetic field}

The strong enhancement of the spin susceptibility $\chi\propto gm$ at
low electron densities can be caused in principle by an increase of
either $g$ or $m$ or both. The effective mass and $g$ factor were
determined separately using the recent theory of
temperature-dependent corrections to conductivity due to
electron-electron interactions \cite{zala01}. Note that its main
advantage compared to the temperature-dependent screening theories
\cite{stern80,gold86,dassarma86,dassarma99} is that spin exchange
effects are treated carefully in the new theory. At intermediate
temperatures, the predicted $\sigma(T)$ is a linear function
\begin{equation}\frac{\sigma(T)}{\sigma_0}=1-A^*k_BT,\qquad\;A^*=-\frac{(1+8F_0^a)gm}{\pi\hbar^2n_s},\label{slope}\end{equation}
where $\sigma_0$ is the value obtained by extrapolating the linear
interval of the $\sigma(T)$ dependence to $T=0$ and the factor of 8
is expected for temperatures lower than the valley splitting in
silicon MOSFETs. The slope, $A^*$, is determined by the Fermi liquid
constants $F_0^a$ and $F_1^s$ that define the renormalization of $g$
and $m$: $g/g_0=1/(1+F_0^a)$ and $m/m_b=1+F_1^s$. Using these
relations one obtains both $g$ and $m$ from the data for the slope
$A^*$ and the product $gm$ \cite{shashkin02b}.

\begin{figure}
\scalebox{0.62}{\includegraphics[clip]{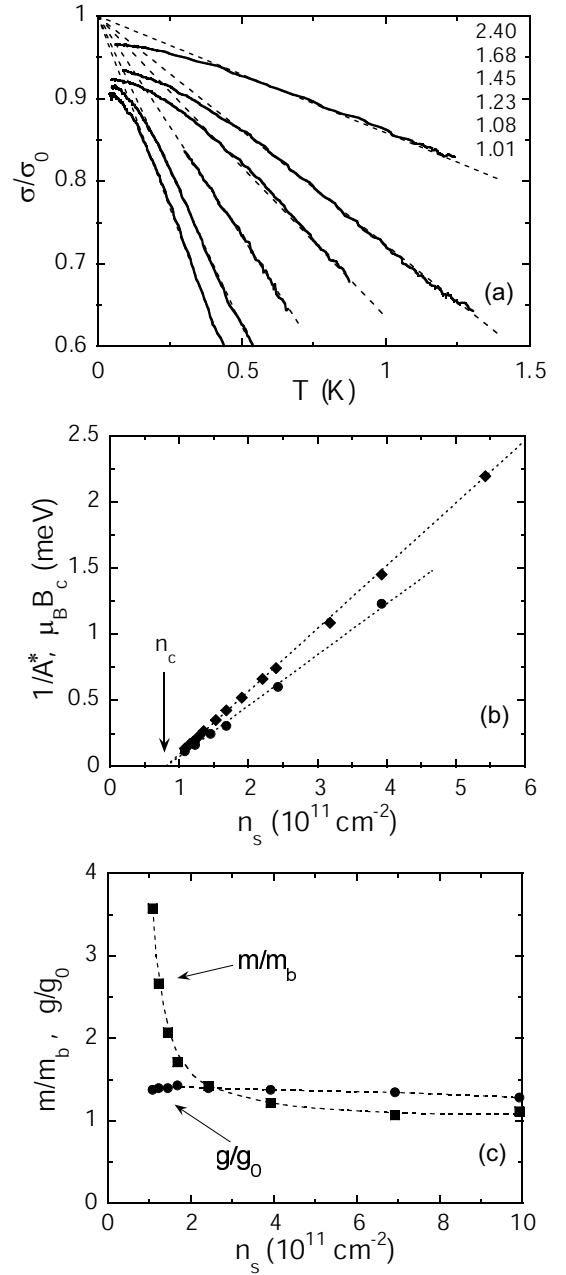}}
\caption{\label{sigma} (a)~The temperature dependence of the
normalized conductivity at different electron densities (indicated in
units of $10^{11}$~cm$^{-2}$) in a low-disordered silicon MOSFET
above the critical electron density for the metal-insulator
transition. The dashed lines are fits to the linear interval of the
dependence. (b)~The inverse slope $1/A^*$ (circles) and the
polarization field $B_c$ (diamonds) as a function of electron
density. The dashed lines are linear fits which extrapolate to the
critical density for the metal-insulator transition. (c)~The
effective mass and $g$ factor versus electron density determined from
an analysis of the temperature-dependent conductivity and
parallel-field magnetoresistance. The dashed lines are guides to the
eye. From Ref.~\cite{shashkin02b}.}
\end{figure}

For sufficiently small deviations $|\sigma/\sigma_0-1|$, the
dependence of the normalized conductivity $\sigma/\sigma_0$ on
temperature at different electron densities above the critical
density $n_c$ for the metal-insulator transition is linear over a
wide enough interval of temperatures (see Fig.~\ref{sigma}(a)). The
inverse slope $1/A^*$ and the value $\mu_BB_c$ are close to each
other in a wide range of electron densities (see
Fig.~\ref{sigma}(b)). Moreover, the low density data for $1/A^*$ are
approximated well by a linear dependence which extrapolates to the
critical density $n_c$ in a way similar to the behavior of $B_c$.
This finding immediately points to approximate constancy of the $g$
factor at low electron densities, according to the functional form of
the slope $A^*$ in Eq.~(\ref{slope}).

Renormalizations $g/g_0$ and $m/m_b$ versus electron density
determined from this analysis corroborate earlier results at high
densities but are striking in the limit of low electron densities
(see Fig.~\ref{sigma}(c)). In the high $n_s$ region, the enhancement
of both $g$ and $m$ is relatively small, both values increasing
slightly with decreasing electron density in agreement with earlier
data \cite{ando82}. Also, the renormalization of the $g$ factor is
dominant compared to that of the effective mass, consistent with
theoretical studies \cite{iwamoto91,kwon94,chen99}. In contrast, in
the low $n_s$ region, the renormalization of the effective mass
increases sharply with decreasing density while the $g$ factor
remains nearly constant. Hence, it is the effective mass, rather than
the $g$ factor, that is responsible for the drastically enhanced spin
susceptibility near the metal-insulator transition.

Normally, no difference is assumed between the interaction parameter
$r_s^*=E_{ee}/E_F$ and the Wigner-Seitz radius $r_s=1/(\pi
n_s)^{1/2}a_B$. The finding of the strongly enhanced effective mass
breaks the equivalence of $r_s^*$ and $r_s$ because they are
connected through the $n_s$ dependent mass: $r_s^*=2(m/m_b)r_s$
(where the factor of 2 comes from the valley degeneracy in silicon
MOSFETs). Therefore, as one approaches the metal-insulator transition
in low-disordered silicon MOSFETs, the interaction parameter $r_s^*$
grows much more rapidly than $r_s$ reaching the values $r_s^*>50$
\cite{shashkin02b}.

In similar experimental verifications of the theory \cite{zala01} on
different 2D carrier systems, much higher values $F_0^a$ than the
expected limit $F_0^a=-1$ for the Stoner instability were found for
the metallic slope of $\sigma(T)$
\cite{proskuryakov02,coleridge02,vitkalov03,pudalov03}. The
relatively small enhancement of the $g$ factor indicates that the
spin exchange effects are not very pronounced. From an experimental
point of view, this raises a problem of comparison between different
theories. Formally, even if there are possible uncertainties in the
theoretical coefficients, both the temperature-dependent screening
theories \cite{stern80,gold86,dassarma86,dassarma99} and the theory
\cite{zala01} describe reasonably well available experimental data
\cite{dassarma03,shashkin03c}. To discriminate between these two,
more detailed comparison with experiment is needed. Note that in all
theories, the slope $A^*$ in Eq.~(\ref{slope}) is proportional to the
effective mass, so the conclusion about the strongly enhanced
effective mass at low electron densities in low-disordered silicon
MOSFETs is basically independent of a particular theory.

\subsubsection{Temperature dependent amplitude of the weak-field Shubnikov-de~Haas oscillations}

\begin{figure}
\scalebox{0.64}{\includegraphics[clip]{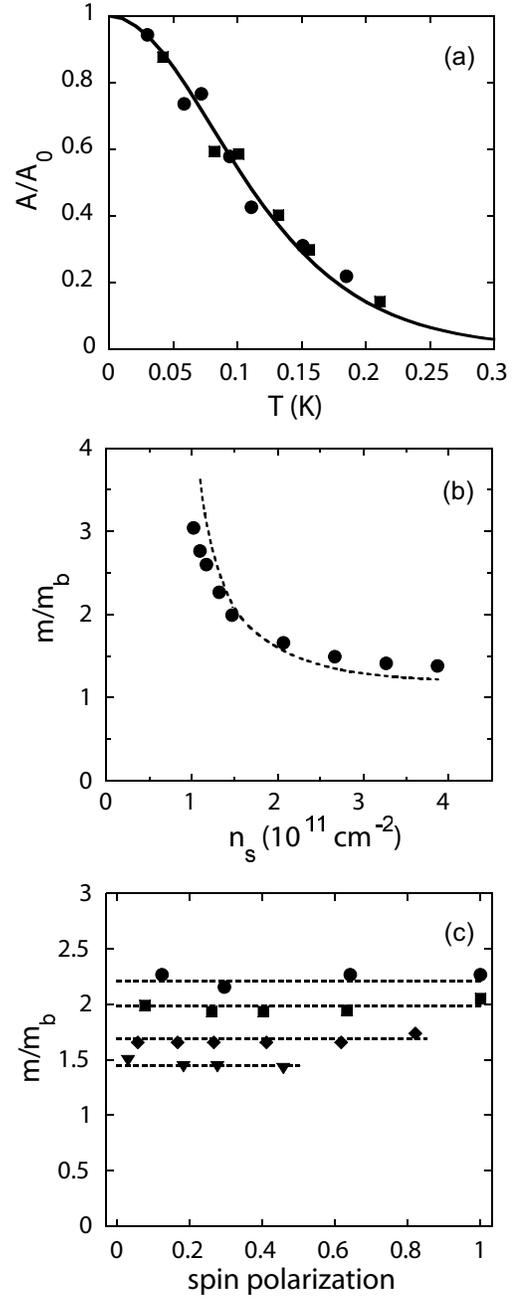}}
\caption{\label{comparison} (a)~Change of the amplitude of the
weak-field Shubnikov-de~Haas oscillations in a low-disordered silicon
MOSFET with temperature at $n_s=1.17\times 10^{11}$~cm$^{-2}$ for
oscillation numbers $\nu=10$ (circles) and $\nu=14$ (squares). The
value of $T$ for the $\nu=10$ data is divided by the factor of 1.4.
The solid line is a fit using Eq.~(\ref{LK}). (b)~Dependence of the
effective mass on electron density determined from an analysis of
Shubnikov-de~Haas oscillations (circles) and from an analysis of
$\sigma(T)$ and $\rho(B_\parallel)$ (dashed line). (c)~The effective
mass versus the degree of spin polarization for the following
electron densities in units of $10^{11}$~cm$^{-2}$: 1.32 (circles),
1.47 (squares), 2.07 (diamonds), and 2.67 (triangles). The dashed
lines are guides to the eye. From Ref.~\cite{shashkin03a}.}
\end{figure}

The claim about the strong increase of the effective mass was
verified based on analysis of the temperature dependence of the
Shubnikov-de~Haas oscillations. The method is similar to that used by
Smith and Stiles \cite{smith72} but is extended to much lower
electron densities and temperatures \cite{shashkin03a}. In the low
temperature limit, the $\rho(T)$ dependence saturates, and the
Lifshitz-Kosevich formula with constant Dingle temperature for the
weak-field oscillation amplitude for the normalized resistance
\begin{eqnarray}\frac{A(T)}{A_0}=\frac{2\pi^2k_BT/\hbar\omega_c}{\sinh(2\pi^2k_BT/\hbar\omega_c)},\nonumber\\A_0=4\exp(-2\pi^2k_BT_D/\hbar\omega_c)\label{LK}\end{eqnarray}
(where $\omega_c=eB_\perp/mc$ is the cyclotron frequency and $T_D$ is
the Dingle temperature) describes damping of the Shubnikov-de~Haas
oscillations with temperature (see Fig.~\ref{comparison}(a)). The
effective mass as a function of electron density determined by this
method agrees well with the data obtained by the procedure described
in the preceding section (see Fig.~\ref{comparison}(b)). The
agreement between the results obtained using two independent methods
supports the validity of both and justifies applicability of
Eq.~(\ref{LK}) to the strongly interacting 2D electron system in
silicon MOSFETs.

To probe a possible contribution from the spin exchange effects to
the effective mass enhancement, a parallel magnetic field component
was introduced to align the electrons' spins. Within the experimental
accuracy, the effective mass does not depend on the degree of spin
polarization $\xi=(B_\perp^2+B_\parallel^2)^{1/2}/B_c$ (see
Fig.~\ref{comparison}(c)). Therefore, the $m(n_s)$ dependence is
robust, and the origin of the mass enhancement has no relation to the
electrons' spins and exchange effects.

\begin{figure}
\scalebox{0.38}{\includegraphics[clip]{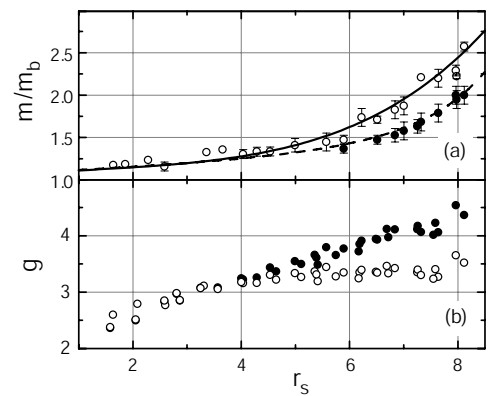}}
\caption{\label{evaluation} The effective mass and $g$ factor in a
dilute silicon MOSFET as a function of $r_s$. Different symbols
correspond to two different assumptions for evaluating $m$ from an
analysis of the high-temperature Shubnikov-de~Haas oscillations: a
temperature-independent $T_D$ (open circles) and a Dingle temperature
that increases linearly with temperature (solid circles). The solid
and dashed lines in (a) are polynomial fits. Adopted from
Ref.~\cite{pudalov02b}.}
\end{figure}

A similar analysis of the Shubnikov-de~Haas oscillations in dilute
silicon MOSFETs at high temperatures $T>0.3$~K, where the low-density
resistivity (and, hence, $T_D$) depends strongly on temperature,
allows an evaluation of the effective mass as well as the $g$ factor
which is calculated from the known value of $gm$ \cite{pudalov02b}
(see Fig.~\ref{evaluation}). The two data sets are obtained based on
an assumption of a temperature-independent Dingle temperature and
that of a Dingle temperature that increases linearly with
temperature. Too large uncertainty of the results makes it impossible
to establish which value (either $g$ or $m$ or both) is responsible
for the strong enhancement of the spin susceptibility. Note that an
attempt to improve the evaluation of the effective mass by justifying
the application of the Lifshitz-Kosevich formula with temperature
dependent $T_D$ \cite{martin03} would lead on the contrary to bigger
deviations of the estimated values of mass from the data obtained in
the low temperature limit (cf. Figs.~\ref{comparison}(b) and
\ref{evaluation}).

An analysis of the temperature dependent amplitude of
Shubnikov-de~Haas oscillations in a dilute 2D electron system in
narrow AlAs quantum wells yielded moderate enhancements of the
effective mass as well as the $g$ factor determined from the known
$gm$ \cite{vakili04}. The observed behavior of $g$ and $m$ is similar
to that found at high electron densities in silicon MOSFETs. This
indicates that the valley origin of the strongly enhanced effective
mass at low electron densities in silicon MOSFETs is not very likely,
even though the lowest accessible densities in narrow AlAs quantum
wells are still too high. Interestingly, the observed values of
$g/g_0$ in the limit of high electron densities in AlAs quantum wells
exceed appreciably the value of $g/g_0=1$ as well as those in silicon
MOSFETs. The increase of the spin susceptibility with the
strain-induced valley polarization, observed at high electron
densities in AlAs quantum wells \cite{shkolnikov04}, is likely to be
connected with the increase in the $g$ factor.

\subsubsection{Spin and cyclotron gaps in strong magnetic fields}

The results for the strong enhancement of the effective mass are also
consistent with the data for spin and cyclotron gaps obtained by
magnetocapacitance spectroscopy. The method is based on determination
of the chemical potential jumps in a 2D electron system when the
filling factor traverses the gaps in the spectrum. A dip in the
magnetocapacitance at integer filling factor is directly related to a
jump of the chemical potential across a corresponding gap in the
spectrum of the 2D electron system \cite{smith85,pudalov86}:
\begin{equation}\frac{1}{C}=\frac{1}{C_0}+\frac{1}{A_ge^2dn_s/d\mu},\label{C}\end{equation}
where $C_0$ is the geometric capacitance and $A_g$ is the sample
area. The chemical potential jump is determined by integrating the
magnetocapacitance over the dip in the low temperature limit where
the magnetocapacitance saturates and becomes independent of
temperature \cite{khrapai03b}. Note that standard measurements of
activation energy yield a mobility gap which may be different from
the gap in the spectrum. This is a serious disadvantage as compared
to the direct method of magnetocapacitance spectroscopy.

\begin{figure}
\scalebox{0.46}{\includegraphics[clip]{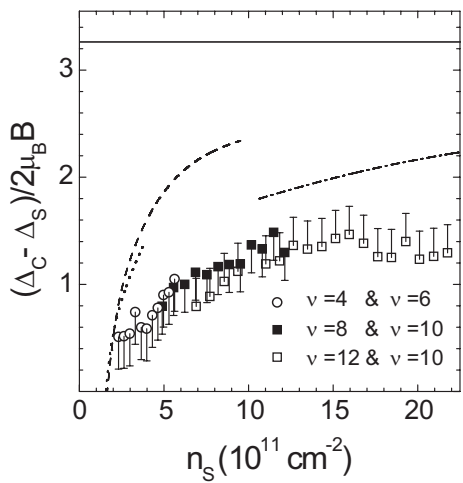}}
\caption{\label{gaps} For silicon MOSFETs, difference of the
normalized values of cyclotron and spin gaps in a perpendicular
magnetic field versus electron density. The level width contribution
is indicated by systematic error bars. Also shown for comparison is
the value ($m_e/m-g$) determined from the data of
Ref.~\cite{shashkin02b} (dashed line), Ref.~\cite{shashkin03a}
(dotted line), and Ref.~\cite{smith72} (dash-dotted line) as well as
using the band electron mass and the $g$ factor in bulk silicon
(solid line). From Ref.~\cite{khrapai03a}.}
\end{figure}

The $g$ factor determined by this method is close to its value in
bulk silicon and does not change with the filling factor
\cite{khrapai03a}, in disagreement with the theory of
exchange-enhanced gaps
\cite{ando74,bychkov81,kallin84,macdonald86,smith92}. The cyclotron
splitting corresponds to the effective mass that is strongly enhanced
at low electron densities (see Fig.~\ref{gaps}). Thus, in strong
magnetic fields, the spin exchange effects are still not pronounced.

It is worth noting that in contrast to the $g$ factor, the valley gap
is strongly enhanced at the lowest filling factors $\nu=1$ and
$\nu=3$ and oscillates with $\nu$ \cite{khrapai03b,khrapai03a}. This
is similar to the behavior of the spin gap in GaAs/AlGaAs
heterostructures \cite{usher90,dolgopolov97}, both of the gaps
increasing linearly with perpendicular magnetic field.

\subsection{Wigner crystal or ferromagnetic Fermi liquid, theoretical approaches}
\label{wigner}

As has been mentioned above, the experimental results obtained in
low-disordered silicon MOSFETs indicate that on the metallic side the
metal-insulator transition is driven by interactions. In contrast, on
the insulating side this is still a classical percolation transition
with no dramatic effects from interactions. One concludes that the
vicinity of either tricritical point on the phase diagram in
Fig.~\ref{diagram} is reached. This is consistent with the fact that
the interaction parameter $r_s^*$ at low electron densities exceeds
the theoretical estimate for the onset of Wigner crystallization,
even though it is not yet clear whether or not electron
crystallization expected in the low density limit is preceded by an
intermediate phase like ferromagnetic Fermi liquid.

To face the problem, two approaches have been formulated. The first
one exploits the Fermi liquid model extending it to relatively large
$r_s^*$. Its outcome is that the renormalization of $g$ is large
compared to that of $m$ \cite{iwamoto91,kwon94,chen99}. In the
limiting case of high $r_s^*$, one may expect a divergence of the $g$
factor that corresponds to the Stoner instability. These predictions
are in contradiction to the experimental data. Firstly, the dilute
system behavior in the regime of the strongly enhanced susceptibility
--- close to the onset of spontaneous spin polarization and Wigner
crystallization --- is governed by the effective mass, rather than
the $g$ factor, through the interaction parameter $r_s^*$. And
secondly, the insensitivity of the effective mass enhancement to spin
exchange effects cannot be accounted for. This discrepancy reduces
somewhat the chances for the occurrence of a ferromagnetic Fermi
liquid that precedes electron crystallization. In principle, should
the spin exchange be small, the spin effects may still come into play
closer to the onset of Wigner crystallization where the Fermi energy
may continue dropping as caused by mass enhancement.

The other theoretical approach is not based on Fermi liquid. In
analogy with He$^3$, it was predicted the existence of an
intermediate phase between the Fermi liquid and the Wigner crystal,
caused by a partial separation of the uniform phases \cite{spivak03}.
It was also predicted that near the crystallization point, the
renormalization of $m$ is dominant compared to that of $g$ and that
the effective mass may diverge at the transition and should increase
with the magnetic field \cite{spivak01}. The strong increase of the
effective mass near electron crystallization also follows from
Gutzwiller's variational method \cite{brinkman70}, which was applied
to silicon MOSFETs \cite{dolgopolov02b}, and from the dynamical
mean-field theory \cite{tanaskovic03}. Although the sharp increase of
the mass is in agreement with the experimental results, the suggested
dependence of $m$ on the degree of spin polarization is not confirmed
by the data.

Concluding this section, I would like to make some more remarks on
the Fermi-liquid-related concepts. An idea was expressed to connect
the observed effective mass enhancement to possible formation of a
coupled-valley state in bivalley electron systems
\cite{dharma03,dharma04}. Still, it is at odds with the fact that
similar results are obtained for single-valley electron systems. An
assumption was made that a plateau at the Fermi energy in the
spectrum $E(k)$ may form leading to a diverging effective mass (see,
e.g., Ref.~\cite{khodel90}). As for now, however, the dependence of
the effective mass on temperature, resulting from the plateau
formation, is not consistent with the experimental findings. A
prediction that the electron density, at which the effective mass
exhibits a sharp increase, is sensitive to disorder \cite{asgari04}
is not confirmed by the experimental data in available samples. After
all, one can simply follow a classical way of introducing
phenomenologically Fermi-liquid parameters as the physical
observables to be determined in experiment.

\section{CONCLUSIONS}

A critical analysis of the available experimental data for 2D
electron systems shows that consequences of the scaling theory of
localization are not confirmed. The main points to be addressed by
theory are the problem of finite bandwidth of the extended states in
the Landau levels, and that of a true metal-insulator transition in
zero magnetic field whose existence is strongly supported in
low-disordered 2D electron systems, but remains uncertain in 2D
electron systems with high disorder. Also, there is still no
theoretical description of the oscillations of the metal-insulator
phase boundary as a function of perpendicular magnetic field.

In the past four years, significant progress was made in
understanding the metallic state in strongly-interacting,
low-disordered 2D electron systems. The state is remarkable by the
strong metallic temperature dependence of the resistivity caused by
electron-electron interaction effects. The spin susceptibility
measured in low-disordered silicon MOSFETs using different
experimental methods shows a sharp increase and possible divergence
at a finite sample-independent electron density $n_\chi$ close to the
critical density $n_c$ for the metal-insulator transition. This
indicates that the metal-insulator transition in clean 2D systems is
driven by interactions. Unlike the Stoner instability, the increase
in the spin susceptibility is caused by the enhanced effective mass
rather than the $g$ factor. The effective mass does not depend on the
degree of spin polarization, so the mass enhancement is not due to
spin exchange. A similar increase in the spin susceptibility is
observed in other 2D carrier systems. It remains to be seen whether
or not it indicates the occurrence of a spontaneous spin polarization
at a finite carrier density.

I am grateful to I.~L. Aleiner, M.~W.~C. Dharma-wardana, V.~T.
Dolgopolov, M.~M. Fogler, V.~F. Gantmakher, D. Heiman, S.~V.
Kravchenko, D.~N. Sheng, and A. Widom for valuable discussions. The
author is supported by the Russian Foundation for Basic Research and
Ministry of Education and Science.


\begin{thebibliography}{apssamp}
\bibitem[1]{wigner34} Wigner E {\it Phys.\ Rev.} {\bf 46} 1002 (1934)
\bibitem[2]{stoner46} Stoner E~C {\it Rep.\ Prog.\ Phys.} {\bf 11} 43
(1946)
\bibitem[3]{landau57} Landau L~D {\it Sov.\ Phys.\ JETP} {\bf 3} 920
(1957)
\bibitem[4]{tanatar89} Tanatar B, Ceperley D~M {\it Phys.\ Rev.\ B}
{\bf 39} 5005 (1989)
\bibitem[5]{attaccalite02} Attaccalite C, Moroni S, Gori-Giorgi P,
Bachelet G~B {\it Phys.\ Rev.\ Lett.} {\bf 88} 256601 (2002)
\bibitem[6]{abrahams79} Abrahams E, Anderson P~W, Licciardello D~C,
Ramakrishnan T~V {\it Phys.\ Rev.\ Lett.} {\bf 42} 673 (1979)
\bibitem[7]{altshuler80} Altshuler B~L, Aronov A~G, Lee P~A {\it Phys.\
Rev.\ Lett.} {\bf 44} 1288 (1980)
\bibitem[8]{finkelstein83} Finkelstein A~M {\it Sov.\ Phys.\ JETP}
{\bf 57} 97 (1983)
\bibitem[9]{finkelstein84} Finkelstein A~M {\it Z.\ Phys.\ B} {\bf 56}
189 (1984)
\bibitem[10]{castellani84} Castellani C, Di Castro C, Lee P~A, Ma M {\it
Phys.\ Rev.\ B} {\bf 30} 527 (1984)
\bibitem[11]{dolan79} Dolan G~J, Osheroff D~D {\it Phys.\ Rev.\ Lett.}
{\bf 43} 721 (1979)
\bibitem[12]{bishop80} Bishop D~J, Tsui D~C, Dynes R~C {\it Phys.\ Rev.\
Lett.} {\bf 44} 1153 (1980)
\bibitem[13]{uren80} Uren M~J, Davies R~A, Pepper M {\it J.\ Phys.\ C}
{\bf 13} L985 (1980)
\bibitem[14]{shashkin93} Shashkin A~A, Kravchenko G~V, Dolgopolov V~T
{\it JETP\ Lett.} {\bf 58} 220 (1993)
\bibitem[15]{shashkin94a} Shashkin A~A, Dolgopolov V~T, Kravchenko G~V
{\it Phys.\ Rev.\ B} {\bf 49} 14486 (1994)
\bibitem[16]{shashkin94b} Shashkin A~A, Dolgopolov V~T, Kravchenko G~V,
Wendel M, Schuster R, Kotthaus J~P, Haug R~J, von Klitzing K, Ploog
K, Nickel H, Schlapp W {\it Phys.\ Rev.\ Lett.} {\bf 73} 3141 (1994)
\bibitem[17]{khmelnitskii84} Khmelnitskii D~E {\it Phys.\ Lett.\ A} {\bf
106} 182 (1984)
\bibitem[18]{laughlin84} Laughlin R~B {\it Phys.\ Rev.\ Lett.} {\bf 52}
2304 (1984)
\bibitem[19]{kravchenko94a} Kravchenko S~V, Kravchenko G~V, Furneaux J~E,
Pudalov V~M, D'Iorio M {\it Phys.\ Rev.\ B} {\bf 50} 8039 (1994)
\bibitem[20]{kravchenko95a} Kravchenko S~V, Mason W~E, Bowker G~E,
Furneaux J~E, Pudalov V~M, D'Iorio M {\it Phys.\ Rev. B} {\bf 51}
7038 (1995)
\bibitem[21]{kravchenko96} Kravchenko S~V, Simonian D, Sarachik M~P,
Mason W, Furneaux J~E {\it Phys.\ Rev.\ Lett.} {\bf 77} 4938 (1996)
\bibitem[22]{kravchenko00a} Kravchenko S~V, Shashkin A~A, Bloore D~A,
Klapwijk T~M {\it Solid\ State\ Commun.} {\bf 116} 495 (2000)
\bibitem[23]{shashkin01a} Shashkin A~A, Kravchenko S~V, Dolgopolov V~T,
Klapwijk T~M {\it Phys.\ Rev.\ Lett.} {\bf 87} 086801 (2001)
\bibitem[24]{vitkalov01a} Vitkalov S~A, Zheng H, Mertes K~M, Sarachik
M~P, Klapwijk T~M {\it Phys.\ Rev.\ Lett.} {\bf 87} 086401 (2001)
\bibitem[25]{shashkin02a} Kravchenko S~V, Shashkin A~A, Dolgopolov V~T
{\it Phys.\ Rev.\ Lett.} {\bf 89} 219701 (2002)
\bibitem[26]{gao02} Gao X~P~A, Mills A~P Jr., Ramirez A~P, Pfeiffer L~N,
West K~W {\it Phys.\ Rev.\ Lett.} {\bf 89} 016801 (2002)
\bibitem[27]{zhu03} Zhu J, Stormer H~L, Pfeiffer L~N, Baldwin K~W,
West K~W {\it Phys.\ Rev.\ Lett.} {\bf 90} 056805 (2003)
\bibitem[28]{shashkin02b} Shashkin A~A, Kravchenko S~V, Dolgopolov V~T,
Klapwijk T~M {\it Phys.\ Rev.\ B} {\bf 66} 073303 (2002)
\bibitem[29]{shashkin03a} Shashkin A~A, Rahimi M, Anissimova S,
Kravchenko S~V, Dolgopolov V~T, Klapwijk T~M {\it Phys.\ Rev.\ Lett.}
{\bf 91} 046403 (2003)
\bibitem[30]{shashkin03b} Shashkin A~A, Kravchenko S~V, Dolgopolov V~T,
Klapwijk T~M {\it J.\ Phys.\ A:\ Math.\ Gen.} {\bf 36} 9237 (2003)
\bibitem[31]{abrahams01} Abrahams E, Kravchenko S~V, Sarachik M~P {\it
Rev.\ Mod.\ Phys.} {\bf 73} 251 (2001)
\bibitem[32]{kravchenko04} Kravchenko S~V, Sarachik M~P {\it Rep.\ Prog.\
Phys.} {\bf 67} 1 (2004)
\bibitem[33]{lozovik75} Lozovik Y~E, Yudson V~I {\it JETP\ Lett.} {\bf
22} 11 (1975)
\bibitem[34]{tsukada77} Tsukada M {\it J.\ Phys.\ Soc.\ Jpn.} {\bf 42}
391 (1977)
\bibitem[35]{maki83} Maki K, Zotos X {\it Phys.\ Rev.\ B} {\bf 28} 4349
(1983)
\bibitem[36]{lam84} Lam P~K, Girvin S~M {\it Phys.\ Rev.\ B} {\bf 30} 473
(1984)
\bibitem[37]{levesque84} Levesque D, Weiss J~J, MacDonald A~M {\it Phys.\
Rev.\ B} {\bf 30} 1056 (1984)
\bibitem[38]{pudalov90} D'Iorio M, Pudalov V~M, Semenchinsky S~G {\it
Phys.\ Lett.\ A} {\bf 150} 422 (1990)
\bibitem[39]{kravchenko91} Kravchenko S~V, Perenboom J~A~A~J, Pudalov V~M
{\it Phys.\ Rev.\ B} {\bf 44} 13513 (1991)
\bibitem[40]{iorio92} D'Iorio M, Pudalov V~M, Semenchinsky S~G {\it
Phys.\ Rev.\ B} {\bf 46} 15992 (1992)
\bibitem[41]{pudalov93a} Pudalov V~M, D'Iorio M, Kravchenko S~V, Campbell
J~W {\it Phys.\ Rev.\ Lett.} {\bf 70} 1866 (1993)
\bibitem[42]{willett88} Willett R~L, Stormer H~L, Tsui D~C, Pfeiffer L~N,
West K~W, Baldwin K~W {\it Phys.\ Rev.\ B} {\bf 38} 7881 (1988)
\bibitem[43]{goldman88} Goldman V~J, Shayegan M, Tsui D~C {\it Phys.\
Rev.\ Lett.} {\bf 61} 881 (1988)
\bibitem[44]{andrei88} Andrei E~Y, Deville G, Glattli D~C, Williams
F~I~B, Paris E, Etienne B {\it Phys.\ Rev.\ Lett.} {\bf 60} 2765
(1988)
\bibitem[45]{willett89} Willett R~L, Stormer H~L, Tsui D~C, Pfeiffer L~N,
West K~W, Shayegan M, Santos M, Sajoto T {\it Phys.\ Rev.\ B} {\bf
40} 6432 (1989)
\bibitem[46]{jiang90} Jiang H~W, Willett R~L, Stormer H~L, Tsui D~C,
Pfeiffer L~N, West K~W {\it Phys.\ Rev.\ Lett.} {\bf 65} 633 (1990)
\bibitem[47]{goldman90} Goldman V~J, Santos M, Shayegan M, Cunningham J~E
{\it Phys.\ Rev.\ Lett.} {\bf 65} 2189 (1990)
\bibitem[48]{williams91} Williams F~I~B, Wright P~A, Clark R~G, Andrei
E~Y, Deville G, Glattli D~C, Probst O, Etienne B, Dorin C, Foxon C~T,
Harris J~J {\it Phys.\ Rev.\ Lett.} {\bf 66} 3285 (1991)
\bibitem[49]{jiang91} Jiang H~W, Stormer H~L, Tsui D~C, Pfeiffer L~N,
West K~W {\it Phys.\ Rev.\ B} {\bf 44} 8107 (1991)
\bibitem[50]{santos92a} Santos M~B, Suen Y~W, Shayegan M, Li Y~P, Engel
L~W, Tsui D~C {\it Phys.\ Rev.\ Lett.} {\bf 68} 1188 (1992)
\bibitem[51]{santos92b} Santos M~B, Jo J, Suen Y~W, Engel L~W, Shayegan M
{\it Phys.\ Rev.\ B} {\bf 46} 13639 (1992)
\bibitem[52]{manoharan94} Manoharan H~C, Shayegan M {\it Phys.\ Rev.\ B}
{\bf 50} 17662 (1994)
\bibitem[53]{klitzing80} von Klitzing K, Dorda G, Pepper M {\it Phys.\
Rev.\ Lett.} {\bf 45} 494 (1980)
\bibitem[54]{kivelson92} Kivelson S~A, Lee D~H, Zhang S~C {\it Phys.\
Rev.\ B} {\bf 46} 2223 (1992)
\bibitem[55]{khmelnitskii92} Khmelnitskii D~E {\it Helv.\ Phys.\ Acta}
{\bf 65} 164 (1992)
\bibitem[56]{huckestein00} Huckestein B {\it Phys.\ Rev.\ Lett.} {\bf 84}
3141 (2000)
\bibitem[57]{dolgopolov92b} Dolgopolov V~T, Kravchenko G~V, Shashkin A~A,
Kravchenko S~V {\it Phys.\ Rev.\ B} {\bf 46} 13303 (1992)
\bibitem[58]{kravchenko95b} Kravchenko S~V, Mason W, Furneaux J~E,
Pudalov V~M {\it Phys.\ Rev.\ Lett.} {\bf 75} 910 (1995)
\bibitem[59]{dultz98} Dultz S~C, Jiang H~W, Schaff W~J {\it Phys.\ Rev.\
B} {\bf 58} R7532 (1998)
\bibitem[60]{hilke00} Hilke M, Shahar D, Song S~H, Tsui D~C, Xie Y~H {\it
Phys.\ Rev.\ B} {\bf 62} 6940 (2000)
\bibitem[61]{glozman95} Glozman I, Johnson C~E, Jiang H~W {\it Phys.\
Rev.\ Lett.} {\bf 74} 594 (1995)
\bibitem[62]{shashkin95} Shashkin A~A, Kravchenko G~V, Dolgopolov V~T,
Kravchenko S~V, Furneaux J~E {\it Phys.\ Rev.\ Lett.} {\bf 75} 2248
(1995)
\bibitem[63]{pudalov93b} Pudalov V~M, D'Iorio M, Campbell J~W {\it JETP\
Lett.} {\bf 57} 608 (1993)
\bibitem[64]{jiang93} Jiang H~W, Johnson C~E, Wang K~L, Hannahs S~T {\it
Phys.\ Rev.\ Lett.} {\bf 71} 1439 (1993)
\bibitem[65]{shahar95a} Shahar D, Tsui D~C, Cunningham J~E {\it Phys.\
Rev.\ B} {\bf 52} R14372 (1995)
\bibitem[66]{hilke97} Hilke M, Shahar D, Song S~H, Tsui D~C, Xie Y~H,
Monroe D {\it Phys.\ Rev.\ B} {\bf 56} R15545 (1997)
\bibitem[67]{sakr01} Sakr M~R, Rahimi M, Kravchenko S~V, Coleridge P~T,
Williams R~L, Lapointe J {\it Phys.\ Rev.\ B} {\bf 64} 161308(R)
(2001)
\bibitem[68]{wang94} Wang T, Clark K~P, Spencer G~F, Mack A~M, Kirk W~P
{\it Phys.\ Rev.\ Lett.} {\bf 72} 709 (1994)
\bibitem[69]{hughes94} Hughes R~J~F, Nicholls J~T, Frost J~E~F, Linfield
E~H, Pepper M, Ford C~J~B, Ritchie D~A, Jones G~A~C, Kogan E, Kaveh M
{\it J.\ Phys.\ Condens.\ Matter} {\bf 6} 4763 (1994)
\bibitem[70]{song97} Song S~H, Shahar D, Tsui D~C, Xie Y~H, Monroe D {\it
Phys.\ Rev.\ Lett.} {\bf 78} 2200 (1997)
\bibitem[71]{lee98} Lee C~H, Chang Y~H, Suen Y~W, Lin H~H {\it Phys.\
Rev.\ B} {\bf 58} 10629 (1998)
\bibitem[72]{hilke98} Hilke M, Shahar D, Song S~H, Tsui D~C, Xie Y~H,
Monroe D {\it Nature} {\bf 395} 675 (1998)
\bibitem[73]{hilke99} Hilke M, Shahar D, Song S~H, Tsui D~C, Xie Y~H,
Shayegan M {\it Europhys.\ Lett.} {\bf 46} 775 (1999)
\bibitem[74]{hanein99} Hanein Y, Nenadovic N, Shahar D, Shtrikman H, Yoon
I, Li C~C, Tsui D~C {\it Nature} {\bf 400} 735 (1999)
\bibitem[75]{popovic97} Popovi\'c D, Fowler A~B, Washburn S {\it Phys.\
Rev.\ Lett.} {\bf 79} 1543 (1997)
\bibitem[76]{coleridge97} Coleridge P~T, Williams R~L, Feng Y, Zawadzki P
{\it Phys.\ Rev.\ B} {\bf 56} R12764 (1997)
\bibitem[77]{shashkin01b} Shashkin A~A, Kravchenko S~V, Klapwijk T~M {\it
Phys.\ Rev.\ Lett.} {\bf 87} 266402 (2001)
\bibitem[78]{papadakis98} Papadakis S~J, Shayegan M {\it Phys.\ Rev.\ B}
{\bf 57} R15068 (1998)
\bibitem[79]{hanein98} Hanein Y, Meirav U, Shahar D, Li C~C, Tsui D~C,
Shtrikman H {\it Phys.\ Rev.\ Lett.} {\bf 80} 1288 (1998)
\bibitem[80]{simmons98} Simmons M~Y, Hamilton A~R, Pepper M, Linfield
E~H, Rose P~D, Ritchie D~A, Savchenko A~K, Griffiths T~G {\it Phys.\
Rev.\ Lett.} {\bf 80} 1292 (1998)
\bibitem[81]{mills99} Mills A~P Jr., Ramirez A~P, Pfeiffer L~N, West K~W
{\it Phys.\ Rev.\ Lett.} {\bf 83} 2805 (1999)
\bibitem[82]{yoon99} Yoon J, Li C~C, Shahar D, Tsui D~C, Shayegan M {\it
Phys.\ Rev.\ Lett.} {\bf 82} 1744 (1999)
\bibitem[83]{simmons00} Simmons M~Y, Hamilton A~R, Pepper M, Linfield
E~H, Rose P~D, Ritchie D~A {\it Phys.\ Rev.\ Lett.} {\bf 84} 2489
(2000)
\bibitem[84]{noh03} Noh H, Lilly M~P, Tsui D~C, Simmons J~A, Hwang E~H,
Das Sarma S, Pfeiffer L~N, West K~W {\it Phys.\ Rev.\ B} {\bf 68}
165308 (2003)
\bibitem[85]{pudalov01} Pudalov V~M, Brunthaler G, Prinz A, Bauer G,
cond-mat/0103087
\bibitem[86]{fogler95} Fogler M~M, Shklovskii B~I {\it Phys.\ Rev.\ B}
{\bf 52} 17366 (1995)
\bibitem[87]{tikofsky00} Tikofsky A~M, Kivelson S~A {\it Phys.\ Rev.\ B}
{\bf 53} R13275 (2000)
\bibitem[88]{ando84} Ando T {\it J.\ Phys.\ Soc.\ Jpn.} {\bf 53} 3126
(1984)
\bibitem[89]{shahbazyan95} Shahbazyan T~V, Raikh M~E {\it Phys.\ Rev.\
Lett.} {\bf 75} 304 (1995)
\bibitem[90]{kagalovsky95} Kagalovsky V, Horovitz B, Avishai Y {\it
Phys.\ Rev.\ B} {\bf 52} R17044 (1995)
\bibitem[91]{gramada96} Gramada A, Raikh M~E {\it Phys.\ Rev.\ B} {\bf
54} 1928 (1996)
\bibitem[92]{haldane97} Haldane F~D~M, Yang K {\it Phys.\ Rev.\ Lett.}
{\bf 78} 298 (1997)
\bibitem[93]{fogler98} Fogler M~M {\it Phys.\ Rev.\ B} {\bf 57} 11947
(1998)
\bibitem[94]{liu96} Liu D~Z, Xie X~C, Niu Q {\it Phys.\ Rev.\ Lett.} {\bf
76} 975 (1996)
\bibitem[95]{xie96} Xie X~C, Liu D~Z, Sundaram B, Niu Q {\it Phys.\ Rev.\
B} {\bf 54} 4966 (1996)
\bibitem[96]{hatsugai99} Hatsugai Y, Ishibashi K, Morita Y {\it Phys.\
Rev.\ Lett.} {\bf 83} 2246 (1999)
\bibitem[97]{yang96} Yang K, Bhatt R~N {\it Phys.\ Rev.\ Lett.} {\bf 76}
1316 (1996)
\bibitem[98]{yang99} Yang K, Bhatt R~N {\it Phys.\ Rev.\ B} {\bf 59} 8144
(1999)
\bibitem[99]{koschny01} Koschny T, Potempa H, Schweitzer L {\it Phys.\
Rev.\ Lett.} {\bf 86} 3863 (2001)
\bibitem[100]{pereira02} Pereira A~L~C, Schulz P~A {\it Phys.\ Rev.\ B}
{\bf 66} 155323 (2002)
\bibitem[101]{koschny03} Koschny T, Schweitzer L {\it Phys.\ Rev.\ B} {\bf
67} 195307 (2003)
\bibitem[102]{sheng97} Sheng D~N, Weng Z~Y {\it Phys.\ Rev.\ Lett.} {\bf
78} 318 (1997)
\bibitem[103]{sheng00} Sheng D~N, Weng Z~Y {\it Phys.\ Rev.\ B} {\bf 62}
15363 (2000)
\bibitem[104]{okamoto95} Okamoto T, Shinohara Y, Kawaji S {\it Phys.\
Rev.\ B} {\bf 52} 11109 (1995)
\bibitem[105]{dultz00} Dultz S~C, Jiang H~W {\it Phys.\ Rev.\ Lett.} {\bf
84} 4689 (2000)
\bibitem[106]{fogler04} Fogler M~M {\it Phys.\ Rev.\ B} {\bf 69} 121409(R)
(2004)
\bibitem[107]{dolgopolov92a} Dolgopolov V~T, Kravchenko G~V, Shashkin A~A
{\it JETP\ Lett.} {\bf 55} 140 (1992)
\bibitem[108]{dolgopolov92c} Dolgopolov V~T, Kravchenko G~V, Shashkin A~A,
Kravchenko S~V {\it JETP\ Lett.} {\bf 55} 733 (1992)
\bibitem[109]{dolgopolov95} Dolgopolov V~T, Shashkin A~A, Kravchenko G~V,
Emeleus C~J, Whall T~E {\it JETP\ Lett.} {\bf 62} 168 (1995)
\bibitem[110]{adkins76} Adkins C~J, Pollitt S, Pepper M {\it J.\ Phys.\ C}
{\bf 37} 343 (1976)
\bibitem[111]{polyakov93} Polyakov D~G, Shklovskii B~I {\it Phys.\ Rev.\
B} {\bf 48} 11167 (1993)
\bibitem[112]{shklovskii84} Shklovskii B~I, Efros A~L {\it Electronic
Properties of Doped Semiconductors} (Springer, New York, 1984)
\bibitem[113]{li94} Li Y~P, PhD thesis, Princeton University, 1994
\bibitem[114]{kukushkin93} Kukushkin I~V, Timofeev V~B {\it Sov.\ Phys.\
Usp.} {\bf 36} 549 (1993)
\bibitem[115]{iordansky82} Iordansky S~V {\it Solid\ State\ Commun.} {\bf
43} 1 (1982)
\bibitem[116]{ando83} Ando T {\it J.\ Phys.\ Soc.\ Jpn.} {\bf 52} 1740
(1983)
\bibitem[117]{aoki83} Aoki H {\it J.\ Phys.\ C} {\bf 16} 1893 (1983)
\bibitem[118]{aoki85} Aoki H, Ando T {\it Phys.\ Rev.\ Lett.} {\bf 54} 831
(1985)
\bibitem[119]{pruisken88} Pruisken A~M~M {\it Phys.\ Rev.\ Lett.} {\bf 61}
1297 (1988)
\bibitem[120]{wei88} Wei H~P, Tsui D~C, Paalanen M~A, Pruisken A~M~M {\it
Phys.\ Rev.\ Lett.} {\bf 61} 1294 (1988)
\bibitem[121]{wakabayashi89} Wakabayashi J, Yamane M, Kawaji S {\it J.\
Phys.\ Soc.\ Jpn.} {\bf 58} 1903 (1989)
\bibitem[122]{koch91a} Koch S, Haug R~J, von Klitzing K, Ploog K {\it
Phys.\ Rev.\ B} {\bf 43} 6828 (1991)
\bibitem[123]{dolgopolov91a} Dolgopolov V~T, Shashkin A~A, Medvedev B~K,
Mokerov V~G {\it Sov.\ Phys.\ JETP} {\bf 72} 113 (1991)
\bibitem[124]{koch91b} Koch S, Haug R~J, von Klitzing K, Ploog K {\it
Phys.\ Rev.\ Lett.} {\bf 67} 883 (1991)
\bibitem[125]{koch92} Koch S, Haug R~J, von Klitzing K, Ploog K {\it
Phys.\ Rev.\ B} {\bf 46} 1596 (1992)
\bibitem[126]{wei92} Wei H~P, Lin S~Y, Tsui D~C, Pruisken A~M~M {\it
Phys.\ Rev.\ B} {\bf 45} 3926 (1992)
\bibitem[127]{hwang93} Hwang S~W, Wei H~P, Engel L~W, Tsui D~C, Pruisken
A~M~M {\it Phys.\ Rev.\ B} {\bf 48} 11416 (1993)
\bibitem[128]{engel93} Engel L~W, Shahar D, Kurdak C, Tsui D~C {\it Phys.\
Rev.\ Lett.} {\bf 71} 2638 (1993)
\bibitem[129]{wei94} Wei H~P, Engel L~W, Tsui D~C {\it Phys.\ Rev.\ B}
{\bf 50} 14609 (1994)
\bibitem[130]{wong95} Wong L~W, Jiang H~W, Trivedi N, Palm E {\it Phys.\
Rev.\ B} {\bf 51} 18033 (1995)
\bibitem[131]{shahar95b} Shahar D, Tsui D~C, Shayegan M, Bhatt R~N,
Cunningham J~E {\it Phys.\ Rev.\ Lett.} {\bf 74} 4511 (1995)
\bibitem[132]{pan97} Pan W, Shahar D, Tsui D~C, Wei H~P, Razeghi M {\it
Phys.\ Rev.\ B} {\bf 55} 15431 (1997)
\bibitem[133]{shahar97} Shahar D, Tsui D~C, Shayegan M, Shimshoni E,
Sondhi S~L {\it Phys.\ Rev.\ Lett.} {\bf 79} 479 (1997)
\bibitem[134]{coleridge99} Coleridge P~T {\it Phys.\ Rev.\ B} {\bf 60}
4493 (1999)
\bibitem[135]{schaijk00} van Schaijk R~T~F, de Visser A, Olsthoorn S~M,
Wei H~P, Pruisken A~M~M {\it Phys.\ Rev.\ Lett.} {\bf 84} 1567 (2000)
\bibitem[136]{dunford00} Dunford R~B, Griffin N, Pepper M, Phillips P~J,
Whall T~E {\it Physica E} {\bf 6} 297 (2000)
\bibitem[137]{dunford01} Dunford R~B, Griffin N, Phillips P~J, Whall T~E
{\it Physica B} {\bf 298} 496 (2001)
\bibitem[138]{hohls01} Hohls F, Zeitler U, Haug R~J {\it Phys.\ Rev.\
Lett.} {\bf 86} 5124 (2001)
\bibitem[139]{hohls02a} Hohls F, Zeitler U, Haug R~J {\it Phys.\ Rev.\
Lett.} {\bf 88} 036802 (2002)
\bibitem[140]{hohls02b} Hohls F, Zeitler U, Haug R~J, Meisels R, Dybko K,
Kuchar F {\it Phys.\ Rev.\ Lett.} {\bf 89} 276801 (2002)
\bibitem[141]{balaban98} Balaban N~Q, Meirav U, Bar-Joseph I {\it Phys.\
Rev.\ Lett.} {\bf 81} 4967 (1998)
\bibitem[142]{shahar98} Shahar D, Hilke M, Li C~C, Tsui D~C, Sondhi S~L,
Cunningham J~E, Razeghi M {\it Solid\ State\ Commun.} {\bf 107} 19
(1998)
\bibitem[143]{arapov02} Arapov Y~G, Alshanskii G~A, Harus G~I, Neverov
V~N, Shelushinina N~G, Yakunin M~V, Kuznetsov O~A {\it
Nanotechnology} {\bf 13} 86 (2002)
\bibitem[144]{huckestein95} Huckestein B {\it Rev.\ Mod.\ Phys.} {\bf 67}
357 (1995)
\bibitem[145]{wysokinski83} Wysokinski K~I, Brenig W {\it Z.\ Phys.\ B}
{\bf 54} 11 (1983)
\bibitem[146]{viehweger90} Viehweger O, Efetov K~B {\it J.\ Phys.\
Condens.\ Matter} {\bf 2} 7049 (1990)
\bibitem[147]{viehweger91} Viehweger O, Efetov K~B {\it Phys.\ Rev.\ B}
{\bf 44} 1168 (1991)
\bibitem[148]{zhang92} Zhang S~C, Kivelson S, Lee D~H {\it Phys.\ Rev.\
Lett.} {\bf 69} 1252 (1992)
\bibitem[149]{wakabayashi88} Wakabayashi J, Fukano A, Kawaji S, Hirakawa
K, Sakaki H, Koike Y, Fukase T {\it J.\ Phys.\ Soc.\ Jpn.} {\bf 57}
3678 (1988)
\bibitem[150]{dorozhkin93} Dorozhkin S~I, Shashkin A~A, Kravchenko G~V,
Dolgopolov V~T, Haug R~J, von Klitzing K, Ploog K {\it JETP\ Lett.}
{\bf 57} 58 (1993)
\bibitem[151]{goldman93} Goldman V~J, Wang J~K, Su B, Shayegan M {\it
Phys.\ Rev.\ Lett.} {\bf 70} 647 (1993)
\bibitem[152]{sajoto93} Sajoto T, Li Y~P, Engel L~W, Tsui D~C, Shayegan M
{\it Phys.\ Rev.\ Lett.} {\bf 70} 2321 (1993)
\bibitem[153]{kravchenko94b} Kravchenko S~V, Furneaux J~E, Pudalov V~M
{\it Phys.\ Rev.\ B} {\bf 49} 2250 (1994)
\bibitem[154]{pudalov94} Pudalov V~M, D'Iorio M, Campbell J~W {\it Surf.\
Sci.} {\bf 305} 107 (1994)
\bibitem[155]{levine83} Levine H, Libby S~B, Pruisken A~M~M {\it Phys.\
Rev.\ Lett.} {\bf 51} 1915 (1983)
\bibitem[156]{khmelnitskii83} Khmelnitskii D~E {\it JETP\ Lett.} {\bf 38}
552 (1983)
\bibitem[157]{dykhne94} Dykhne A~M, Ruzin I~M {\it Phys.\ Rev.\ B} {\bf
50} 2369 (1994)
\bibitem[158]{ruzin95} Ruzin I, Feng S {\it Phys.\ Rev.\ Lett.} {\bf 74}
154 (1995)
\bibitem[159]{burgess00} Burgess C~P, Dib R, Dolan B~P {\it Phys.\ Rev.\
B} {\bf 62} 15359 (2000)
\bibitem[160]{wei86} Wei H~P, Tsui D~C, Pruisken A~M~M {\it Phys.\ Rev.\
B} {\bf 33} 1488 (1986)
\bibitem[161]{halperin82} Halperin B~I {\it Phys.\ Rev.\ B} {\bf 25} 2185
(1982)
\bibitem[162]{QHE87} {\it The Quantum Hall Effect} Ed. by Prange R~E,
Girvin S~M (Springer-Verlag, 1987)
\bibitem[163]{buttiker88} B\"uttiker M {\it Phys.\ Rev.\ B} {\bf 38} 9375
(1988)
\bibitem[164]{laughlin81} Laughlin R~B {\it Phys.\ Rev.\ B} {\bf 23} 5632
(1981)
\bibitem[165]{widom82} Widom A, Clark T~D {\it J.\ Phys.\ D} {\bf 15} L181
(1982)
\bibitem[166]{dolgopolov90} Dolgopolov V~T, Zhitenev N~B, Shashkin A~A
{\it JETP\ Lett.} {\bf 52} 196 (1990)
\bibitem[167]{dolgopolov91b} Dolgopolov V~T, Zhitenev N~B, Shashkin A~A
{\it Europhys.\ Lett.} {\bf 14} 255 (1991)
\bibitem[168]{dolgopolov92d} Dolgopolov V~T, Shashkin A~A, Zhitenev N~B,
Dorozhkin S~I, von Klitzing K {\it Phys.\ Rev.\ B} {\bf 46} 12560
(1992)
\bibitem[169]{dolgopolov93} Dolgopolov V~T, Shashkin A~A, Kravchenko G~V,
Dorozhkin S~I, von Klitzing K {\it Phys.\ Rev.\ B} {\bf 48} 8480
(1993)
\bibitem[170]{jeanneret95} Jeanneret B, Hall B~D, Buhlmann H~J, Houdre R,
Ilegems M, Jeckelmann B, Feller U {\it Phys.\ Rev.\ B} {\bf 51} 9752
(1995)
\bibitem[171]{watts98} Watts J~P, Usher A, Matthews A~J, Zhu M, Elliott M,
Herrenden-Harker W~G, Morris P~R, Simmons M~Y, Ritchie D~A {\it
Phys.\ Rev.\ Lett.} {\bf 81} 4220 (1998)
\bibitem[172]{honold99} Honold M~M, Harrison N, Singleton J, Nam M~S,
Blundell S~J, Mielke C~H, Kartsovnik M~V, Kushch N~D {\it Phys.\
Rev.\ B} {\bf 59} R10417 (1999)
\bibitem[173]{dolgopolov01} Dolgopolov V~T, Shashkin A~A, Broto J~M,
Rakoto H, Askenazy S {\it Phys.\ Rev.\ Lett.} {\bf 86} 5566 (2001)
\bibitem[174]{dolgopolov91c} Dolgopolov V~T, Kravchenko G~V, Shashkin A~A
{\it Solid\ State\ Commun.} {\bf 78} 999 (1991)
\bibitem[175]{haug93} Haug R~J {\it Semicond.\ Sci.\ Technol.} {\bf 8} 131
(1993)
\bibitem[176]{fontein91} Fontein P~F, Kleinen J~A, Hendriks P, Blom F~A~P,
Wolter J~H, Lochs H~G~M, Driessen F~A~J~M, Giling L~J, Beenakker
C~W~J {\it Phys.\ Rev.\ B} {\bf 43} 12090 (1991)
\bibitem[177]{kent92} Kent A~J, McKitterick D~J, Challis L~J, Hawker P,
Mellor C~J, Henini M {\it Phys.\ Rev.\ Lett.} {\bf 69} 1684 (1992)
\bibitem[178]{merz93} Merz R, Keilmann F, Haug R~J, Ploog K {\it Phys.\
Rev.\ Lett.} {\bf 70} 651 (1993)
\bibitem[179]{shashkin94c} Shashkin A~A, Kent A~J, Harrison P~A,
Strickland K~R, Eaves L, Henini M {\it Semicond.\ Sci.\ Technol.}
{\bf 9} 2110 (1994)
\bibitem[180]{shashkin94d} Shashkin A~A, Kent A~J, Harrison P~A, Eaves L,
Henini M {\it Phys.\ Rev.\ B} {\bf 49} 5379 (1994)
\bibitem[181]{haren95a} van Haren R~J~F, Blom F~A~P, Wolter J~H {\it
Phys.\ Rev.\ Lett.} {\bf 74} 1198 (1995)
\bibitem[182]{haren95b} van Haren R~J~F, de Lange W, Blom F~A~P, Wolter
J~H {\it Phys.\ Rev.\ B} {\bf 52} 5760 (1995)
\bibitem[183]{shashkin97} Shashkin A~A, Kent A~J, Owers-Bradley J~R, Cross
A~J, Hawker P, Henini M {\it Phys.\ Rev.\ Lett.} {\bf 79} 5114 (1997)
\bibitem[184]{wei98} Wei Y~Y, Weis J, von Klitzing K, Eberl K {\it Phys.\
Rev.\ Lett.} {\bf 81} 1674 (1998)
\bibitem[185]{tessmer98} Tessmer S~H, Glicofridis P~I, Ashoori R~C,
Levitov L~S, Melloch M~R {\it Nature} {\bf 392} 51 (1998)
\bibitem[186]{mccormick99} McCormick K~L, Woodside M~T, Huang M, Wu M,
McEuen P~L, Duruoz C, Harris J~S Jr. {\it Phys.\ Rev.\ B} {\bf 59}
4654 (1999)
\bibitem[187]{yacoby99} Yacoby A, Hess H~F, Fulton T~A, Pfeiffer L~N, West
K~W {\it Solid\ State\ Commun.} {\bf 111} 1 (1999)
\bibitem[188]{shashkin99} Shashkin A~A, Kent A~J, Hawker P, Henini M {\it
Phys.\ Rev.\ B} {\bf 60} R16307 (1999)
\bibitem[189]{zhitenev00} Zhitenev N~B, Fulton T~A, Yacoby A, Hess H~F,
Pfeiffer L~N, West K~W {\it Nature} {\bf 404} 473 (2000)
\bibitem[190]{finkelstein00a} Finkelstein G, Glicofridis P~I, Ashoori R~C,
Shayegan M {\it Science} {\bf 289} 90 (2000)
\bibitem[191]{finkelstein00b} Finkelstein G, Glicofridis P~I, Tessmer S~H,
Ashoori R~C, Melloch M~R {\it Phys.\ Rev.\ B} {\bf 61} R16323 (2000)
\bibitem[192]{woodside01} Woodside M~T, Vale C, McEuen P~L, Kadow C,
Maranowski K~D, Gossard A~C {\it Phys.\ Rev.\ B} {\bf 64} 041310(R)
(2001)
\bibitem[193]{zalinge01} van Zalinge H, \"Ozyilmaz B, B\"ohm A, van der
Heijden R~W, Wolter J~H, Wyder P {\it Phys.\ Rev.\ B} {\bf 64} 235303
(2001)
\bibitem[194]{glicofridis02} Glicofridis P~I, Finkelstein G, Ashoori R~C,
Shayegan M {\it Phys.\ Rev.\ B} {\bf 65} 121312(R) (2002)
\bibitem[195]{chklovskii92} Chklovskii D~B, Shklovskii B~I, Glazman L~I
{\it Phys.\ Rev.\ B} {\bf 46} 4026 (1992)
\bibitem[196]{ebert85} Ebert G, von Klitzing K, Weimann G {\it J.\ Phys.\
C} {\bf 18} L257 (1985)
\bibitem[197]{pudalov85} Pudalov V~M, Semenchinskii S~G {\it JETP\ Lett.}
{\bf 42} 232 (1985)
\bibitem[198]{shashkin86} Shashkin A~A, Dolgopolov V~T, Dorozhkin S~I {\it
Sov.\ Phys.\ JETP} {\bf 64} 1124 (1986)
\bibitem[199]{wiegers99} Wiegers S~A~J, Lok J~G~S, Jeuken M, Zeitler U,
Maan J~C, Henini M {\it Phys.\ Rev.\ B} {\bf 59} 7323 (1999)
\bibitem[200]{sarachik99} Sarachik M~P, Kravchenko S~V {\it Proc.\ Natl.\
Acad.\ Sci.\ USA} {\bf 96} 5900 (1999)
\bibitem[201]{kravchenko00b} Kravchenko S~V, Klapwijk T~M {\it Phys.\
Rev.\ Lett.} {\bf 84} 2909 (2000)
\bibitem[202]{jaroszynski02} Jaroszy\'nski J, Popovi\'c D, Klapwijk T~M
{\it Phys.\ Rev.\ Lett.} {\bf 89} 276401 (2002)
\bibitem[203]{simonian97a} Simonian D, Kravchenko S~V, Sarachik M~P {\it
Phys.\ Rev.\ B} {\bf 55} R13421 (1997)
\bibitem[204]{okamoto99} Okamoto T, Hosoya K, Kawaji S, Yagi A {\it Phys.\
Rev.\ Lett.} {\bf 82} 3875 (1999)
\bibitem[205]{vitkalov00} Vitkalov S~A, Zheng H, Mertes K~M, Sarachik M~P,
Klapwijk T~M {\it Phys.\ Rev.\ Lett.} {\bf 85} 2164 (2000)
\bibitem[206]{vitkalov01b} Vitkalov S~A, Sarachik M~P, Klapwijk T~M {\it
Phys.\ Rev.\ B} {\bf 64} 073101 (2001)
\bibitem[207]{bogdanovich02} Bogdanovich S, Popovi\'c D {\it Phys.\ Rev.\
Lett.} {\bf 88} 236401 (2002)
\bibitem[208]{leturcq03} Leturcq R, L'Hote D, Tourbot R, Mellor C~J,
Henini M {\it Phys.\ Rev.\ Lett.} {\bf 90} 076402 (2003)
\bibitem[209]{stern80} Stern F {\it Phys.\ Rev.\ Lett.} {\bf 44} 1469
(1980)
\bibitem[210]{gold86} Gold A, Dolgopolov V~T {\it Phys.\ Rev.\ B} {\bf 33}
1076 (1986)
\bibitem[211]{dassarma86} Das Sarma S {\it Phys.\ Rev.\ B} {\bf 33} 5401
(1986)
\bibitem[212]{dassarma99} Das Sarma S, Hwang E~H {\it Phys.\ Rev.\ Lett.}
{\bf 83} 164 (1999)
\bibitem[213]{fang68} Fang F~F, Stiles P~J {\it Phys.\ Rev.} {\bf 174} 823
(1968)
\bibitem[214]{smith72} Smith J~L, Stiles P~J {\it Phys.\ Rev.\ Lett.} {\bf
29} 102 (1972)
\bibitem[215]{ando74} Ando T, Uemura Y {\it J.\ Phys.\ Soc.\ Jpn.} {\bf
37} 1044 (1974)
\bibitem[216]{bychkov81} Bychkov Yu~A, Iordanskii S~V, Eliashberg G~M {\it
JETP\ Lett.} {\bf 33} 143 (1981)
\bibitem[217]{kallin84} Kallin C, Halperin B~I {\it Phys.\ Rev.\ B} {\bf
30} 5655 (1984)
\bibitem[218]{macdonald86} MacDonald A~H, Oji H~C~A, Liu K~L {\it Phys.\
Rev.\ B} {\bf 34} 2681 (1986)
\bibitem[219]{smith92} Smith A~P, MacDonald A~H, Gumbs G {\it Phys.\ Rev.\
B} {\bf 45} 8829 (1992)
\bibitem[220]{simonian97b} Simonian D, Kravchenko S~V, Sarachik M~P,
Pudalov V~M {\it Phys.\ Rev.\ Lett.} {\bf 79} 2304 (1997)
\bibitem[221]{pudalov97} Pudalov V~M, Brunthaler G, Prinz A, Bauer G {\it
JETP\ Lett.} {\bf 65} 932 (1997)
\bibitem[222]{pudalov02a} Pudalov V~M, Brunthaler G, Prinz A, Bauer G {\it
Phys.\ Rev.\ Lett.} {\bf 88} 076401 (2002)
\bibitem[223]{dolgopolov00} Dolgopolov V~T, Gold A {\it JETP\ Lett.} {\bf
71} 27 (2000)
\bibitem[224]{pudalov02b} Pudalov V~M, Gershenson M~E, Kojima H, Butch N,
Dizhur E~M, Brunthaler G, Prinz A, Bauer G {\it Phys.\ Rev.\ Lett.}
{\bf 88} 196404 (2002)
\bibitem[225]{vitkalov02} Vitkalov S~A, Sarachik M~P, Klapwijk T~M {\it
Phys.\ Rev.\ B} {\bf 65} 201106(R) (2002)
\bibitem[226]{sarachik03} Sarachik M~P, Vitkalov S~A {\it J.\ Phys.\ Soc.\
Jpn.} {\bf 72} 57 (2003)
\bibitem[227]{prus03} Prus O, Yaish Y, Reznikov M, Sivan U, Pudalov V {\it
Phys.\ Rev.\ B} {\bf 67} 205407 (2003)
\bibitem[228]{dolgopolov02a} Dolgopolov V~T, Gold A {\it Phys.\ Rev.\
Lett.} {\bf 89} 129701 (2002)
\bibitem[229]{gold02} Gold A, Dolgopolov V~T {\it J.\ Phys.:\ Condens.\
Matter} {\bf 14} 7091 (2002)
\bibitem[230]{batke86} Batke E, Tu C~W {\it Phys.\ Rev.\ B} {\bf 34} 3027
(1986)
\bibitem[231]{sarma00} Das Sarma S, Hwang E~H {\it Phys.\ Rev.\ Lett.}
{\bf 84} 5596 (2000)
\bibitem[232]{tutuc02} Tutuc E, Melinte S, Shayegan M {\it Phys.\ Rev.\
Lett.} {\bf 88} 036805 (2002)
\bibitem[233]{noh02} Noh H, Lilly M~P, Tsui D~C, Simmons J~A, Hwang E~H,
Das Sarma S, Pfeiffer L~N, West K~W {\it Phys.\ Rev.\ B} {\bf 68}
165308 (2002)
\bibitem[234]{tutuc03} Tutuc E, Melinte S, De Poortere E~P, Shayegan M,
Winkler R {\it Phys.\ Rev.\ B} {\bf 67} 241309(R) (2003)
\bibitem[235]{vakili04} Vakili K, Shkolnikov Y~P, Tutuc E, De Poortere
E~P, Shayegan M {\it Phys.\ Rev.\ Lett.} {\bf 92} 226401 (2004)
\bibitem[236]{dolgopolov04} Dolgopolov V~T, Deviatov E~V, Shashkin A~A,
Wieser U, Kunze U, Abstreiter G, Brunner K {\it Superlattices\
Microstruct.} {\bf 33} 271 (2003)
\bibitem[237]{okamoto04} Okamoto T, Ooya M, Hosoya K, Kawaji S {\it Phys.\
Rev.\ B} {\bf 69} 041202 (2004)
\bibitem[238]{zala01} Zala G, Narozhny B~N, Aleiner I~L {\it Phys.\ Rev.\
B} {\bf 64} 214204 (2001)
\bibitem[239]{ando82} Ando T, Fowler A~B, Stern F {\it Rev.\ Mod.\ Phys.}
{\bf 54} 437 (1982)
\bibitem[240]{iwamoto91} Iwamoto N {\it Phys.\ Rev.\ B} {\bf 43} 2174
(1991)
\bibitem[241]{kwon94} Kwon Y, Ceperley D~M, Martin R~M {\it Phys.\ Rev.\
B} {\bf 50} 1684 (1994)
\bibitem[242]{chen99} Chen G~H, Raikh M~E {\it Phys.\ Rev.\ B} {\bf 60}
4826 (1999)
\bibitem[243]{proskuryakov02} Proskuryakov Y~Y, Savchenko A~K, Safonov
S~S, Pepper M, Simmons M~Y, Ritchie D~A {\it Phys.\ Rev.\ Lett.} {\bf
89} 076406 (2002)
\bibitem[244]{coleridge02} Coleridge P~T, Sachrajda A~S, Zawadzki P {\it
Phys.\ Rev.\ B} {\bf 65} 125328 (2002)
\bibitem[245]{vitkalov03} Vitkalov S~A, James K, Narozhny B~N, Sarachik
M~P, Klapwijk T~M {\it Phys.\ Rev.\ B} {\bf 67} 113310 (2003)
\bibitem[246]{pudalov03} Pudalov V~M, Gershenson M~E, Kojima H, Brunthaler
G, Prinz A, Bauer G {\it Phys.\ Rev.\ Lett.} {\bf 91} 126403 (2003)
\bibitem[247]{dassarma03} Das Sarma S, Hwang E~H {\it Phys.\ Rev.\ Lett.}
{\bf 93} 269703 (2004)
\bibitem[248]{shashkin03c} Shashkin A~A, Dolgopolov V~T, Kravchenko S~V
{\it Phys.\ Rev.\ Lett.} {\bf 93} 269705 (2004)
\bibitem[249]{martin03} Martin G~W, Maslov D~L, Reizer M~Yu {\it Phys.\
Rev.\ B} {\bf 68} 241309(R) (2003)
\bibitem[250]{shkolnikov04} Shkolnikov Y~P, Vakili K, De Poortere E~P,
Shayegan M {\it Phys.\ Rev.\ Lett.} {\bf 92} 246804 (2004)
\bibitem[251]{smith85} Smith T~P, Goldberg B~B, Stiles P~J, Heiblum M {\it
Phys.\ Rev.\ B} {\bf 32} 2696 (1985)
\bibitem[252]{pudalov86} Pudalov V~M, Semenchinskii S~G {\it JETP\ Lett.}
{\bf 44} 677 (1986)
\bibitem[253]{khrapai03b} Khrapai V~S, Shashkin A~A, Dolgopolov V~T {\it
Phys.\ Rev.\ B} {\bf 67} 113305 (2003)
\bibitem[254]{khrapai03a} Khrapai V~S, Shashkin A~A, Dolgopolov V~T {\it
Phys.\ Rev.\ Lett.} {\bf 91} 126404 (2003)
\bibitem[255]{usher90} Usher A, Nicholas R~J, Harris J~J, Foxon C~T {\it
Phys.\ Rev.\ B} {\bf 41} 1129 (1990)
\bibitem[256]{dolgopolov97} Dolgopolov V~T, Shashkin A~A, Aristov A~V,
Schmerek D, Hansen W, Kotthaus J~P, Holland M {\it Phys.\ Rev.\
Lett.} {\bf 79} 729 (1997)
\bibitem[257]{spivak03} Spivak B {\it Phys.\ Rev.\ B} {\bf 67} 125205
(2003)
\bibitem[258]{spivak01} Spivak B {\it Phys.\ Rev.\ B} {\bf 64} 085317
(2001)
\bibitem[259]{brinkman70} Brinkman W~F, Rice T~M {\it Phys.\ Rev.\ B} {\bf
2} 4302 (1970)
\bibitem[260]{dolgopolov02b} Dolgopolov V~T {\it JETP\ Lett.} {\bf 76} 377
(2002)
\bibitem[261]{tanaskovic03} Tanaskovi\'c D, Dobrosavljevi\'c V, Abrahams
E, Kotliar G {\it Phys.\ Rev.\ Lett.} {\bf 91} 066603 (2003)
\bibitem[262]{dharma03} Dharma-wardana M~W~C {\it Europhys.\ Lett.} {\bf
67} 552 (2004)
\bibitem[263]{dharma04} Dharma-wardana M~W~C, Perrot F {\it Phys.\ Rev.\
B} {\bf 70} 035308 (2004)
\bibitem[264]{khodel90} Khodel V~A, Shaginyan V~R {\it JETP\ Lett.} {\bf
51} 553 (1990)
\bibitem[265]{asgari04} Asgari R, Davoudi B, Tanatar B {\it Solid\ State\
Commun.} {\bf 130} 13 (2004)
\end{thebibliography}
\end{document}